\documentclass[a4paper,12pt]{article}

\usepackage{amssymb}                                 %
\usepackage{amsmath}                                 %
\newcommand{\be}{\begin{equation}}                                     %
\newcommand{\ee}{\end{equation}}                                       %
\newcommand{\bea}{\begin{eqnarray}}                                    %
\newcommand{\eea}{\end{eqnarray}}                                      %
\newcommand{\ba}{\begin{align}}                                        %
\newcommand{\ea}{\end{align}}                                          %
\renewcommand{\theequation}{\arabic{section}.\arabic{equation}}        %
\newcommand{\ri}{\mathrm{i}}               %
\newcommand{\rr}{\mathrm{r}}               %
\newcommand{\dd}{\mathrm{d}}               %
\newcommand{\ad}{\mathrm{ad}}              %
\newcommand{\half}{\frac{1}{2}}            %
\newcommand{\quarter}{\frac{1}{4}}         %
\newcommand{\bR}{\mathbb{R}}               %
\newcommand{\bC}{\mathbb{C}}               %
\newcommand{\bZ}{\mathbb{Z}}               %
\newcommand{\cD}{\mathcal{D}}              %
\newcommand{\cH}{\mathcal{H}}              %
\newcommand{\cU}{\mathcal{U}}              %
\newcommand{\cB}{\mathcal{B}}              %
\newcommand{\cT}{\mathcal{T}}              %
\newcommand{\cC}{\mathcal{C}}              %
\newcommand{\cL}{\mathcal{L}}              %
\newcommand{\cF}{\mathcal{F}}              %
\newcommand{\cA}{\mathcal{A}}              %
\newcommand{\cK}{\mathcal{K}}              %
\newcommand{\cG}{\mathcal{G}}              %
\newcommand{\cJ}{\mathcal{J}}              %
\newcommand{\cV}{\mathcal{V}}              %
\newcommand{\1}{\mathrm{\bf{1}}}           %
\title{\Large{\textbf{Twisted spin Sutherland models
from quantum Hamiltonian reduction}}}

\author{\normalsize{L.~FEH\'ER${}^a$ and B.G.~PUSZTAI${}^b$}}

\date{}

\begin{document}
\maketitle
\begin{center}
${}^a$Department of Theoretical Physics, MTA  KFKI RMKI\\
1525 Budapest 114, P.O.B. 49,  Hungary, and\\
Department of Theoretical Physics, University of Szeged\\
Tisza Lajos krt 84-86, H-6720 Szeged, Hungary\\
e-mail: \texttt{lfeher@rmki.kfki.hu}

\bigskip
${}^b$Centre de recherches math\'ematiques, Universit\'e de Montr\'eal\\
C.P. 6128, succ. centre ville, Montr\'eal, Qu\'ebec, Canada H3C 3J7, and\\
Department of Mathematics and Statistics, Concordia University\\
1455 de Maisonneuve Blvd. West, Montr\'eal, Qu\'ebec, Canada H3G 1M8\\
e-mail: \texttt{pusztai@CRM.UMontreal.CA}
\end{center}

\bigskip

\begin{abstract}
Recent general results on Hamiltonian reductions
under polar group actions are applied to study
some reductions of the free particle governed
by the Laplace-Beltrami operator
of a compact, connected, simple Lie group.
The reduced systems
associated with arbitrary finite dimensional irreducible representations of the group
by using the symmetry induced by twisted
conjugations are described in detail.
These systems generically yield integrable Sutherland type
many-body models with spin, which are called
twisted spin Sutherland models
if the underlying twisted conjugations
are built on non-trivial Dynkin diagram automorphisms.
The spectra of these models can be calculated, in principle,
by solving certain Clebsch-Gordan problems, and
 the result is presented for the models associated
with the symmetric tensorial powers of the
defining representation of $SU(N)$.
\end{abstract}

\newpage

\section{Introduction}
\setcounter{equation}{0}

The investigation of one-dimensional
integrable many-body systems initiated by Calogero \cite{Cal}, Sutherland \cite{Sut}
and others is still an actively pursued field of mathematical physics.
These models possess interesting physical applications and are
closely related to harmonic analysis and to the theory of special functions.
See e.g.~the reviews in \cite{OPRep, Heck, Dunkl, SutWSci, Cher,Sas, Eting, Poly}.

The aim of this paper is to apply the quantum Hamiltonian reduction
approach developed in \cite{FP_polar, FP_ROMP}  under certain general assumptions
to construct and analyze new examples
of spin Sutherland type models, where by `Sutherland type' we mean that
the interaction potential involves the function $1/\sin^{2}x$ in association with a root system.
A first impression about our models may  be obtained by viewing the special cases
in equations (\ref{5.19}) and (\ref{5.20}) below, where the operators
in the numerators act on  internal `spin' degrees of freedom and the presence of
$1/\cos^2x$ in the interaction indicates the twisted character of these
Sutherland type models.
These examples illustrate the statement \cite{FP_polar} that quantum Hamiltonian
reductions of the free particle on a Lie group or on a symmetric space under polar
actions of compact symmetry groups lead to Calogero-Sutherland type models
with internal degrees of freedom in general.

A polar action of a compact Lie group, $G$,  is an isometric action
on a complete Riemannian manifold,
$Y$, which permits the introduction of adapted polar coordinates
 as defined in \cite{PT} systematizing  classical examples.
In the cases of our interest the radial coordinates run over a suitable
Abelian Lie group, whose Lie algebra carries an associated root system.
The quantum Hamiltonian reduction amounts to restricting the Laplace-Beltrami operator,
$\Delta_Y$, of $Y$  to vector valued generalized spherical functions,
which are wave functions on $Y$ belonging to some fixed representation type
under the
symmetry group $G$. If this representation is the trivial one, then the reduction
yields the radial part of the Laplace-Beltrami operator.

It was a pioneering observation of Olshanetsky and Perelomov \cite{OP} that
the radial part of the Laplace-Beltrami operator of any Riemannian symmetric space
provides the Hamiltonian of a Calogero-Sutherland type model at special
coupling constants.
One may also notice  by inspecting examples that
if one considers spherical functions corresponding to an arbitrary representation of the symmetry
group, then the angular part of the Laplace-Beltrami
operator contributes an interaction term of spin Calogero-Sutherland type
for general families of polar actions (e.g., for the so-called Hermann actions
recalled in Chapter 3).
We say that  `spin' degrees of freedom are present
to express the fact that the reduced
wave functions are vector valued in all but some exceptional cases.
It is also important to note that the reduced
wave functions are actually scalar valued for certain
non-trivial representation types under some symmetry groups.
This was pointed out by Etingof, Frenkel and Kirillov in \cite{EFK}, where they
used this observation  to derive
the standard Sutherland model with arbitrary integer coupling constants
from Hamiltonian reduction of the free particle on the group manifold $SU(N)$.

In standard harmonic analysis \cite{Helg} the spherical functions and
the spectrum of the Laplace-Beltrami operator are among the central
objects of interest.
As discussed above, it is well-known that several powerful results of harmonic analysis
can be translated into statements about many-body models with spin defined
by Hamiltonian reduction.
However, it appears  that this idea  has not yet been systematically exploited.
In fact, since the standard harmonic analysis
approach has the limitation of giving spinless models only at special coupling
constants,  the attention was mainly focused on the more algebraic methods
that are capable to overcome this limitation, such as the techniques
relying on  Dunkl operators and Hecke algebras (reviewed in
\cite{Heck,Dunkl,Cher, Poly}).
Our opinion  is that although the standard harmonic analysis approach
gives indeed only a limited class of spinless many-body models, it would be worth to
develop this point of view systematically, partly since the resulting
models with spin are interesting, and
partly since it is still not clear what is  the full set of  spinless
models that can be described in this framework.
For recent studies concerning the spinless  models, see \cite{Obl,FP_LMP}.

The program to develop the classical and quantum Hamiltonian reduction approach to
spin Calogero-Sutherland type models in general terms and to explore
the set of systems that it covers was advanced in the  recent papers
\cite{Hoch,FP1,FP2,FP_polar}.
In this paper we deal with certain novel examples at the quantum mechanical level,
which we call twisted spin Sutherland models since they result from quantum
Hamiltonian reduction based on the so-called twisted conjugation action of a compact simple
Lie group on itself.
The definition of this polar action can be seen in
equation (\ref{3.12}) below, where $\Theta$ is an automorphism of the compact
symmetry group $G$.
It generalizes the ordinary conjugation action, which is recovered if $\Theta$ is the
identity automorphism.
The geometry of twisted conjugations has been recently investigated in \cite{W,M,MW},
and we shall use some results of these references.
Our work  builds also on  \cite{FP1}, where we have described
the classical mechanical counterparts of the twisted spin Sutherland models.

The content of the present paper and our main results can be outlined as follows.
In Chapter 2 we review the quantum Hamiltonian reduction
of the free particle focusing mainly on the description of the
reduced systems obtained with the aid of polar actions of compact
Lie groups. This chapter is based on our earlier work \cite{FP_polar, FP_ROMP}
(see also \cite{IT}).
Chapter 3 contains the derivation of the spin Sutherland models
associated with involutive Dynkin diagram automorphisms of the
simple Lie algebras.
The models are displayed in Proposition 3.1, which is a new result.
The twisted spin Sutherland models correspond to non-trivial automorphisms, but
the previously studied case (e.g.~\cite{EFK}) of the trivial automorphism is also covered.
Chapter 4 is devoted to explaining how the diagonalization of the spin
Sutherland Hamiltonians of Proposition 3.1 can be performed in principle
by solving certain Clebsch-Gordan
problems in the representation theory of the underlying symmetry group $G$.
The construction of the models involves choosing a representation of $G$,
and in Chapter 5 we analyze the examples associated with the symmetric
tensorial powers of the defining representation of $G=SU(N)$.
We first present these models, in Propositions 5.1 and 5.2,
by using the realization of the symmetric tensors
in terms of $N$ harmonic oscillators.
We then determine the spectrum of the Hamiltonian in these cases
by applying some standard results (the so-called Pieri formulae)
of representation theory.  The spectra of these twisted spin Sutherland
Hamiltonians are given by Theorem 5.3, which is one of our
main results.
This generalizes the well-known formula (\ref{5.33}) of the spectrum
of the standard spinless Sutherland model (\ref{5.30}), here  recovered
from Hamiltonian reduction, at integer couplings, as a warm-up exercise \cite{EFK}.
Finally, our conclusions and comments on open problems are collected in Chapter 6, and
the three appendices contain some technical details.

In what follows we make an effort to present the analysis in a self-contained manner,
as an application of a general framework that can be used in future works as well.

\newpage
\section{Some facts about quantum Hamiltonian reduction}
\setcounter{equation}{0}

We here summarize basic facts about quantum Hamiltonian
reductions of free particles on complete Riemannian manifolds under
isometric actions of compact Lie groups. The free Hamiltonian
will be taken to be the scalar Laplace-Beltrami operator, and we
shall assume the existence of generalized polar coordinates adapted
to the group action. No new results are contained in this chapter,
which mainly serves to fix notations for the subsequent
developments. Proofs and more details can be found in
\cite{FP_polar, FP_ROMP, IT}.

\subsection{Definitions}

Suppose that a compact Lie group $G$ acts on a complete, connected, smooth
Riemannian manifold $(Y, \eta)$ in an isometric manner.
This means that we are given a smooth left-action
\be
\phi \colon G \times Y \rightarrow Y,
\quad
(g, y) \mapsto \phi(g, y) = \phi_g(y) = g . y
\label{2.1}\ee
of $G$ on $Y$ satisfying $\phi_g^* \eta = \eta$ for every $g \in G$.
Then the measure $\mu_Y$ on $Y$ induced by the metric $\eta$ is
$G$-invariant, and the natural action of $G$ on the Hilbert space
$L^2(Y, \dd \mu_Y)$ gives rise to a  continuous unitary representation
of $G$,
\be
U \colon G \rightarrow \cU(L^2(Y, \dd \mu_Y)),
\quad
g \mapsto U(g).
\label{2.2}\ee
Obviously,
this unitary representation of $G$ commutes with the restriction,
$\Delta_Y^0$, of the Laplace-Beltrami operator, $\Delta_Y$, to the space
of smooth complex functions of compact support, $C_c^\infty(Y)$,
\be
\Delta_Y^0 U(g) f = U(g) \Delta_Y^0 f
\quad
\forall g \in G, \forall f \in C_c^\infty(Y).
\label{2.3}\ee
The  Laplace-Beltrami
operator with domain $C_c^\infty(Y)$,
\be
\Delta_Y^0 := \Delta_Y|_{C_c^\infty(Y)} \colon C_c^\infty(Y) \rightarrow C_c^\infty(Y),
\label{2.4}\ee
 is an essentially self-adjoint operator
on $L^2(Y, \dd \mu_Y)$, i.e., its closure $\bar \Delta_Y^0$ is self-adjoint (see e.g.~\cite{Lands}).
As a result, the pair $(L^2(Y, \dd \mu_Y), -\frac{1}{2}\bar \Delta_Y^0)$ is a quantum
mechanical system with symmetry group $G$, which is
a quantum mechanical analogue of the free classical point mass
moving along  geodesics on the Riemannian manifold $(Y, \eta)$.

Let $\rho: G \to \cU(V_\rho)$ be a continuous unitary irreducible
representation (finite dimensional `irrep')
of $G$, and denote
the corresponding complex conjugate representation by $(\rho^*, V_{\rho^*})$.
It is not difficult to exhibit the following unitary equivalence of $G$-representations:
\be
L^2(Y,\dd \mu_Y) \cong \oplus_\rho\, L^2(Y, V_\rho, \dd \mu_Y)^G \otimes  V_{\rho^*}
\label{2.5}\ee
where the sum is over  the  pairwise inequivalent irreps and
$L^2(Y, V_\rho, \dd \mu_Y)^G$ is the space of $G$-singlets in the Hilbert space of
$V_\rho$-valued square integrable functions on $Y$.
(Of course, $L^2(Y, V_\rho, \dd \mu_Y)\cong L^2(Y, \dd\mu_Y) \otimes V_\rho$
as a representation space of $G$, and its scalar product uses also the scalar product
 on $V_\rho$.)
In terms of the decomposition (2.5) of $L^2(Y,\dd \mu_Y)$,
the action of $G$ is non-trivial only on the factors $V_{\rho^*}$, and the action of
$\bar \Delta_Y^0$ is non-trivial only on the `multiplicity spaces'
$L^2(Y, V_\rho, \dd \mu_Y)^G$.
By keeping only these multiplicity spaces, one obtains a reduced  quantum system
for every $G$ irrep $(\rho, V_\rho)$.

To be more explicit, the reduced Hilbert space
consists of the $V_\rho$-valued $G$-equivariant
square-integrable functions on $Y$,
\be
L^2(Y, V_\rho, \dd \mu_Y)^G
:= \{ f \, | \, f \in L^2(Y, V_\rho, \dd \mu_Y), f \circ \phi_g = \rho(g) \circ f
\quad \forall g \in G \}.
\label{2.6}\ee
The Laplace-Beltrami operator acts naturally on the space of the $V_\rho$-valued
$G$-equivariant smooth functions of compact support, simply componentwise.
This gives the operator
\be
\Delta_\rho \colon C_c^\infty(Y, V_\rho)^G \rightarrow C_c^\infty(Y, V_\rho)^G,
\label{2.7}\ee
which is  essentially self-adjoint on the Hilbert space $L^2(Y, V_\rho, \dd \mu_Y)^G$.
The closure of $\Delta_\rho$ is the Hamiltonian of the reduced quantum system,
\be
(L^2(Y, V_\rho, \dd \mu_Y)^G, -\frac{1}{2}\bar \Delta_\rho),
\label{2.8}\ee
that arises from the free particle in association with the irrep $(\rho, V_\rho)$.

Although the reduced quantum system is already completely fixed
by (2.8), it is desirable (e.g., for the physical interpretation)
 to realize
the reduced state-space $L^2(Y, V_\rho, \dd \mu_Y)^G$ as a Hilbert space of appropriate functions
on
the reduced configuration space $Y_{\mathrm{red}} := Y / G$, and
the reduced Hamiltonian operator as a differential
operator over this space.
Here one encounters a difficulty since
the orbit space $Y / G$ is not a smooth
manifold but a \emph{stratified space} in general, i.e., a disjoint union
of countably many smooth Riemannian manifolds.
However, restricting to the generic points forming the submanifold
$\check{Y} \subset Y$ of \emph{principal orbit type}, one obtains a smooth fiber
bundle $\pi \colon \check{Y} \rightarrow \check{Y} / G$,
and the smooth part of the reduced configuration space,
$\check{Y}_{\mathrm{red}} := \check{Y} / G$, can be endowed with a
reduced Riemannian metric, $\eta_{\mathrm{red}}$, in such a way
that $\pi$ becomes a Riemannian
submersion\footnote{For detailed discussions on the principal orbit type and
stratifications, we recommend the references \cite{Davis,GOV}.}.
The crucial facts are that  $\check Y$ is dense and open in $Y$
and its complement is of zero measure.
Since the wave functions belonging to the domain of essential self-adjointness
$C_c^\infty(Y, V_\rho)^G$ (2.7)  can be
recovered from their restriction to $\check Y$,
the above mentioned difficulty is only apparent.
It is possible to
work out a complete characterization of the reduced systems in terms
of the smooth reduced configuration manifold
$(\check{Y}_{\mathrm{red}}, \eta_{\mathrm{red}})$ in general.
Next we present this
under a simplifying assumption that
holds in the examples of our interest.

\subsection{Characterization of the reduced systems}

From now on we assume that $G$ acts on $(Y, \eta)$ in a \emph{polar}
manner, that is, the action $\phi$ (\ref{2.1}) admits \emph{sections}
in the sense of Palais and Terng \cite{PT}.
Recall that a section $\Sigma \subset Y$ is a connected, closed, regularly
embedded smooth submanifold of $Y$ that meets every $G$-orbit and it does
so orthogonally at every intersection point of $\Sigma$ with an
orbit. By its embedding, $\Sigma$ inherits a Riemannian metric $\eta_\Sigma$,
which induces a measure $\mu_\Sigma$ on $\Sigma$.
For a section $\Sigma$, denote by $\check{\Sigma}$ a
connected component of the manifold $\hat{\Sigma} := \check{Y} \cap \Sigma$.
The isotropy subgroups of all elements of $\hat{\Sigma}$ are the same
and for a fixed section we define $K := G_y$ for $y \in \hat{\Sigma}$.
The restriction of
$\pi \colon \check{Y} \rightarrow \check{Y} / G$
onto $\check{\Sigma}$ provides an isometric diffeomorphism between the
Riemannian manifolds $(\check{\Sigma}, \eta_{\check{\Sigma}})$ and
$(\check{Y}_{\mathrm{red}}, \eta_{\mathrm{red}})$, and the $G$-equivariant
diffeomorphism
\be
\check{\Sigma} \times (G / K) \ni (q, g K) \mapsto \phi_g(q) \in \check{Y}
\label{2.9}\ee
defines a global trivialization of the fiber bundle
$\pi \colon \check{Y} \rightarrow \check{Y} / G$.
We below recall the characterization of the reduced systems in terms of
the reduced configuration space $(\check{\Sigma}, \eta_{\check{\Sigma}}) \cong
(\check{Y}_{\mathrm{red}}, \eta_{\mathrm{red}})$.

First, we introduce the $\bC$-linear space
\be
\mathrm{Fun}(\check \Sigma, V_\rho^K):= \{ f\in C^\infty(\check\Sigma, V_\rho^K)\,\vert\,
\exists \cF\in C_c^\infty(Y, V_\rho)^G, \,\, f= \cF\vert_{\check \Sigma}\,\},
\label{2.10}\ee
where $V_\rho^K$ is the subspace of \emph{$K$-invariant vectors}
in the representation space $V_\rho$. We assume that $\dim(V_\rho^K) > 0$.
As a vector space, $\mathrm{Fun}(\check{\Sigma}, V_\rho^K)$ is naturally
isomorphic to $C_c^\infty(Y, V_\rho)^G$ and can be equipped with
a scalar product induced by this isomorphism.
Using also that $C_c^\infty(Y, V_\rho)^G$ is
dense in $L^2(Y, V_\rho, \dd \mu_Y)^G$, for the closure of
$\mathrm{Fun}(\check{\Sigma}, V_\rho^K)$ we obtain the Hilbert space isomorphism
$\mathrm{\overline{Fun}}(\check{\Sigma}, V_\rho^K) \cong L^2(Y, V_\rho, \dd \mu_Y)^G$.

We also introduce the  density function,
$\delta \colon \check{\Sigma} \rightarrow (0, \infty)$,
as follows.
The $G$-orbit $G . q \subset Y$ through any point $q \in \check{\Sigma}$ is an
embedded submanifold of $Y$ and by its embedding it inherits
a Riemannian metric, $\eta_{G . g}$.
We let
\be
\delta(q) := \mbox{volume of the Riemannian manifold } (G . q, \eta_{G . q}),
\label{2.11}\ee
where, of course, the volume is understood with respect the measure belonging
to $\eta_{G . q}$.

Now let us consider the Lie algebra $\cG := \mathrm{Lie}(G)$ and its
subalgebra $\cK := \mathrm{Lie}(K)$.  Choose  a $G$-invariant
positive definite scalar product, $\cB$, on $\cG$,
which gives rise to the
orthogonal decomposition
\be
\cG = \cK \oplus \cK^\perp.
\label{2.12}\ee
For any $\xi \in \cG$ denote by $\xi^\sharp$ the corresponding  vector
field on $Y$.
At each  point $q \in \check{\Sigma}$,
the linear map
\be
\cK^\perp \ni \xi \mapsto \xi_q^\sharp \in T_q Y
\label{2.13}\ee
is injective and permits to define the `inertia operator'
$\cJ(q) \in GL(\cK^\perp)$  by requiring
\be
\eta_q(\xi^\sharp_q, \zeta^\sharp_q) = \cB(\xi, \cJ(q) \zeta)
\quad
\forall \xi, \zeta \in \cK^\perp.
\label{2.14}\ee
Note that $\cJ(q)$ is symmetric and positive definite
with respect to the restriction of the scalar product $\cB$ to
$\cK^\perp$.
In $\cK^\perp$ we introduce dual bases $\{ T_\alpha \}$ and  $\{
T^\alpha \}$,  $\cB(T^\alpha, T_\beta) =
\delta^\alpha_\beta$, and we let
\be
b_{\alpha, \beta}(q) :=
\cB(T_\alpha, \cJ(q) T_\beta),
\qquad
b^{\alpha, \beta}(q) :=
\cB(T^\alpha, \cJ(q)^{-1} T^\beta).
\label{2.15}\ee
The matrix
$b^{\alpha, \beta}(q)$ is the inverse of $b_{\alpha, \beta}(q)$, and
for the density function we have
\be \delta(q) = C
\sqrt{\vert \det(b_{\alpha, \beta}(q))\vert}
 \qquad \forall q \in \check{\Sigma},
\label{2.16}\ee
where $C>0$ is some constant.

Finally, for the Lie algebra representation belonging to the unitary
representation $(\rho, V_\rho)$ of $G$, we introduce the notation
$\rho': \cG \to u(V_\rho)$,
where $u(V_\rho)$ is the Lie algebra of anti-hermitian operators on $V_\rho$.
The following result is then proved in \cite{FP_polar} (see also \cite{FP_ROMP}).

\bigskip
\noindent
\textbf{Proposition 2.1.} \emph{Suppose that $\phi$ (\ref{2.1}) is a polar action
of the compact Lie group $G$ on the Riemannian manifold $(Y,\eta)$ and
choose a section $\Sigma$ for this action.
Then, using the notations introduced above,
the reduced system (2.8) associated
with a continuous unitary irreducible representation
$(\rho, V_\rho)$ of  $G$ can be identified with the pair
$(L^2(\check{\Sigma}, V_\rho^K, \dd \mu_{\check{\Sigma}}),-\frac{1}{2}\bar \Delta_{\mathrm{red}})$,
where the reduced Laplace-Beltrami operator
\be
\Delta_{\mathrm{red}}
:= \Delta_{\check{\Sigma}} - \delta^{-\frac{1}{2}} \Delta_{\check{\Sigma}}(\delta^{\frac{1}{2}})
+ b^{\alpha, \beta} \rho'(T_\alpha) \rho'(T_\beta)
\label{2.17}\ee
with domain $\delta^{\frac{1}{2}} \mathrm{Fun}(\check{\Sigma}, V_\rho^K)$
is essentially self-adjoint on  the Hilbert space
$L^2(\check{\Sigma}, V_\rho^K, \dd \mu_{\check{\Sigma}})$,
and  $\bar \Delta_{\mathrm{red}}$ denotes  its  self-adjoint closure.
}

\bigskip
\noindent
{\bf Remark 2.2.}
Proposition 2.1 utilizes   the identification
$L^2(Y,V_\rho, \dd \mu_Y)^G\cong
L^2(\check{\Sigma}, V_\rho^K, \delta \dd \mu_{\check{\Sigma}})$
(obtained by restricting the $G$-equivariant wave functions to $\check \Sigma$)
as well as the isometry between
$L^2(\check{\Sigma}, V_\rho^K, \delta \dd \mu_{\check{\Sigma}})$
and $L^2(\check{\Sigma}, V_\rho^K, \dd \mu_{\check{\Sigma}})$
defined by multiplying the restricted wave functions
 by $\delta^{\frac{1}{2}}$.
The latter  step is natural since the measure
$\mu_{\check{\Sigma}}$ is directly defined by the reduced Riemannian
metric $\eta_{\check \Sigma}$ that enters also $\Delta_{\check \Sigma}$.

\bigskip
\noindent
{\bf Remark 2.3.}
One can view the reduced systems of Proposition 2.1 from an alternative perspective
that sheds light on a generalization of the well-known Weyl group invariance
of the standard Calogero-Sutherland models.
The essential point is that the
restriction of the elements of $C_c^\infty(Y, V_\rho)^G$ to $\check \Sigma$
can be implemented in two steps, initially restricting them to $\hat \Sigma = \check Y \cap \Sigma$.
The wave functions obtained in this first step are equivariant
with respect to the residual symmetry transformations
generated by the elements of $G$ that map $\Sigma$ (or equivalently $\hat \Sigma$)
to itself. More precisely, since $K$ acts trivially on $\Sigma$,
these transformations form  the factor group
\be
W:= N_G(\Sigma)/K,
\label{2.18}\ee
where $N_G(\Sigma)$ contains the $\Sigma$-preserving elements of $G$.
It is proved in \cite{PT} that $W$ is a finite group for any
polar action.
The representation of $G$ on $V_\rho$ induces a representation of
$W$ on $V^K_\rho$,
and $W$  permutes the connected components of $\hat \Sigma$ by its action.
We now have  the following natural Hilbert space isomorphisms:
\be
L^2(Y,V_\rho, \dd \mu_Y)^G\cong L^2(\check{\Sigma}, V_\rho^K, \dd \mu_{\check{\Sigma}}) \cong
L^2(\hat{\Sigma}, V_\rho^K, \frac{1}{\vert W\vert} \dd \mu_{\hat{\Sigma}})^W
\cong L^2(\Sigma, V_\rho^K, \frac{1}{\vert W\vert} \dd \mu_{\Sigma})^W,
\label{2.19}\ee
where the last equality is based on the fact that  $\Sigma \setminus \hat \Sigma$ has measure zero.
The Hilbert space
$L^2(\hat{\Sigma}, V_\rho^K, \frac{1}{\vert W\vert} \dd \mu_{\hat{\Sigma}})^W$
carries the Hamiltonian given by the same formula as
$\Delta_{\mathrm{red}}$ (\ref{2.17}) but using $\hat \Sigma$ instead of $\check \Sigma$.
This operator is essentially self-adjoint on the domain
$\delta^{\frac{1}{2}} \mathrm{Fun}(\hat{\Sigma}, V_\rho^K)$, where $\delta^{\frac{1}{2}}$
is defined on $\hat \Sigma$ as in (\ref{2.11}).
The space $\mathrm{Fun}(\hat \Sigma, V_\rho^K)$, defined similarly to (\ref{2.10}),
consists of $W$-equivariant functions.
We shall further elaborate this remark, with full proofs, elsewhere.

\section{Construction of twisted spin Sutherland  models}
\setcounter{equation}{0}

Next we briefly present a general Lie theoretic framework that
permits to construct a large family of spin Sutherland type models.
Then we describe some examples  (which we call twisted spin Sutherland models) in detail;
the full family will be studied in a future publication.

Take  $Y$ to be a compact, connected, semisimple Lie group endowed with a
biinvariant Riemannian metric $\eta$ induced by a multiple of the
Killing form. Let $G$ be an arbitrary \emph{symmetric subgroup}
of the product Lie group $Y \times Y$, that is, $G$ satisfies
\be
(Y \times Y)^\sigma_0 \subset G \subset (Y \times Y)^\sigma,
\label{3.1}\ee
where $\sigma$ is an involutive automorphism of $Y \times Y$,
$(Y \times Y)^\sigma$ denotes the fixed point subgroup of $\sigma$,
and $(Y \times Y)^\sigma_0$ is the connected component of
the identity in $(Y \times Y)^\sigma$.
In this general case the following  action (often called `Hermann action')
\be
\phi \colon G \times Y \rightarrow Y,
\quad
((a, b), y) \mapsto \phi_{(a, b)}(y) := a y b^{-1}
\label{3.2}\ee
of $G$ on $Y$ is known to be \emph{hyperpolar}, which means that this action
is polar in such a way that the sections are \emph{flat} in the
induced metric.
In fact,
the sections $\Sigma \subset Y$ are provided by certain tori of $Y$
associated with Abelian subalgebras
of the correct dimension lying in the subspace $(T_e(G . e))^\perp$ of
$T_e Y$.
These results, and many more on related matters, can be found in \cite{HPTT,Kol}.

For example,
choose an arbitrary automorphism $\Theta \in \mathrm{Aut}(Y)$ and set
\be
\sigma(y_1, y_2) := (\Theta^{-1}(y_2),
\Theta(y_1)) \quad \forall (y_1, y_2) \in Y \times Y.
\label{3.3}\ee
By projection to the second factor,
the symmetric subgroup
\be
G := \{ (\Theta^{-1}(g), g) \, | \, g \in Y \} \subset Y \times Y
\label{3.4}\ee
can be identified with $Y$, $G \cong Y$,
and  (\ref{3.2}) then becomes the
action of $Y$ on itself by \emph{$\Theta$-twisted conjugations}.
Indeed, after identifying $G$ with $Y$, equation (\ref{3.2}) yields the action whereby
$g\in Y$ sends $y\in Y$ to $\Theta^{-1}(g)y g^{-1}$, which is ordinary conjugation
by $g$ if $\Theta$ is the identity.
In the most interesting cases $\Theta$ corresponds to a Dynkin diagram symmetry of $Y$.
Some of the resulting spin Sutherland models have been investigated in our earlier
work \cite{FP1} at the classical level.
After fixing the necessary group theoretical conventions,
we describe the quantum mechanical counterparts of these models in subsection 3.2.
For simplicity, in what follows we assume
that the underlying Lie group is simple and  simply connected.

\subsection{Conventions}

Let $G$ be a compact, connected, simply-connected, simple  Lie group with
 fixed maximal torus $T \subset G$.  Set $r:= \mathrm{dim}(T)$.
Denote by  $\cA$ and $\cH$ the complexifications of the real
Lie algebras  $\cG:= \mathrm{Lie}(G)$ and
$\cT := \mathrm{Lie}(T)$.
Then $\cA$ is a complex simple
Lie algebra with  Cartan subalgebra  $\cH$,
and we choose a
polarization $\Phi = \Phi_+ \cup \Phi_-$ for the root system $\Phi$
of $(\cH, \cA)$ and a set of simple roots
$\{ \varphi_k \}_{k = 1}^r \subset \Phi_+$.
We also select root vectors $\{ X_\varphi \}_{\varphi \in \Phi}$
satisfying
\be
\langle X_\varphi, X_{-\varphi} \rangle = 1
\quad
\forall \varphi \in \Phi_+,
\label{3.5}\ee
where $\langle \, , \rangle \colon \cA \times \cA \rightarrow \bC$ is
a convenient positive multiple of the Killing form of $\cA$.  We let
\be
T_{\varphi_k} := [ X_{\varphi_k}, X_{-\varphi_k} ] \in \cH,
\quad
1 \leq k \leq r,
\label{3.6}\ee
and thus we have
\be
\cT = \ri \cH_\rr
\quad\hbox{with}\quad
\cH_\rr := \mathrm{span}_{\bR}\{ T_{\varphi_k} \, | \, 1 \leq k \leq r \}.
\label{3.7}\ee
By a suitable choice of the root vectors we may assume that
\be
\cG =
\cT \oplus \left( \oplus_{\varphi \in \Phi_+} \bR Y_\varphi \right)
\oplus \left( \oplus_{\varphi \in \Phi_+} \bR Z_\varphi \right),
\label{3.8}\ee
where
\be
Y_\varphi :=\frac{\ri}{\sqrt{2}}(X_\varphi + X_{-\varphi})
\quad \mbox{and} \quad
Z_\varphi :=\frac{1}{\sqrt{2}}(X_\varphi - X_{-\varphi})
\quad
\forall \varphi \in \Phi_+.
\label{3.9}\ee
The bilinear form
\be
\cB := -\langle \, , \rangle|_{\cG \times \cG} \colon \cG \times \cG \rightarrow \bR
\label{3.10}\ee
is a $G$-invariant positive definite  scalar product on
$\cG$ and  (\ref{3.9}) defines an
orthonormal set of vectors with respect to $\cB$.
We equip $G$ with the biinvariant Riemannian metric $\eta$ induced by
this scalar product.

Any symmetry $\theta$ of the Dynkin diagram of $\cA$ extends to
an automorphism of $\cA$ by the requirement
\be
\theta(X_{\pm\varphi_k}) = X_{\pm\theta(\varphi_k)}
\quad 1 \leq k \leq r.
\label{3.11}\ee
The resulting automorphism $\theta \in \mathrm{Aut}(\cA)$ preserves
the real algebras $\cG$ and $\cT$ and gives rise  to an automorphism $\Theta$ of the group
$G$, which  maps the torus  $T$ to itself.
Then  $G$ acts on itself by  the $\Theta$-twisted conjugations mentioned before.
We here designate this action as
\be
I^\Theta \colon
G \times G \rightarrow G,
\quad (g, y) \mapsto I^\Theta_g(y) :=
\Theta^{-1}(g) y g^{-1}.
\label{3.12}\ee
A \emph{section} for this hyperpolar action
is furnished \cite{MW} by the fixed point subgroup  $T^\Theta$
of $\Theta$ in the maximal torus $T$.
The isotropy subgroup of the elements of principal $I^\Theta$
orbit type  in $T^\Theta$ is given by $T^\Theta$ itself.
In the notations used in Proposition 2.1, we have
\be
\Sigma =T^\Theta,
\quad
K=T^\Theta.
\label{3.13}\ee
Correspondingly,
$\check \Sigma=\check T^{\Theta}$ stands below for a connected component of the set of
$I^\Theta$-regular elements in $T^\Theta$.
Note also that in the fixed
point subalgebra $\cT^\theta$ of $\theta$ in $\cT$ one can
choose a bounded open domain
 $\check{\cT}^\theta$ (a generalized Weyl alcove) such that the exponential map restricted to
$\check{\cT}^\theta$ provides a one-to-one parametrization of
$\check{T}^\Theta$.
Following \cite{MW}, we explain in Appendix B that
$\check \cT^\theta$  can be characterized as the interior of a fundamental domain
for the action of a `twisted affine Weyl group' on $\cT^\theta$.
In the above we used that $G$ is not only connected but also simply connected, since
otherwise $T^\Theta$ would not be always connected \cite{M,MW}, and here we wish to avoid this
complication.

From now on we assume that the automorphism (\ref{3.11}) is  \emph{involutive},
and denote by
$\cA^\pm$, $\cG^\pm$, $\cH^\pm$, $\cH_\rr^\pm$ and $\cT^\pm$ the
corresponding eigensubspaces of $\theta$ with eigenvalues $\pm 1$.
Then $\cA^+$ is a complex simple Lie algebra with
Cartan subalgebra $\cH^+$, and $\cA^-$ is an irreducible module
of $\cA^+$ whose non-zero weights have \emph{multiplicity one} \cite{Kac}.
Moreover,
$(\cT^+, \cG^+)$ is a compact real form of $(\cH^+, \cA^+)$ with
the associated real irreducible module $\cG^-$. Based on the standard
`folding procedure', detailed for example in \cite{FP1},  one can construct
convenient bases
for all these spaces from the above Weyl-Chevalley basis of $\cA$.
We next display the bases needed later.

Let ${\mathfrak{R}}$ be the set of roots of $(\cH^+, \cA^+)$ and
let ${\mathfrak{W}}$ be the set of non-zero weights for $(\cH^+, \cA^-)$.
We choose
root vectors $X^+_\alpha$ $(\alpha \in {\mathfrak{R}})$
and weight vectors $X^-_\lambda$ $(\lambda \in {\mathfrak{W}})$
normalized by
\be
\langle X^+_\alpha, X^+_{-\alpha} \rangle = 1
\quad
\forall \alpha \in {\mathfrak{R}},
\qquad
\langle X^-_\lambda, X^-_{-\lambda} \rangle = 1
\quad
\forall \lambda \in {\mathfrak{W}}.
\label{3.14}\ee
Selecting positive roots and weights,
${\mathfrak{R}} = {\mathfrak{R}}_+ \cup {\mathfrak{R}}_-$ and
${\mathfrak{W}} = {\mathfrak{W}}_+ \cup {\mathfrak{W}}_-$,
we define
\bea
&& Y^+_\alpha :=\frac{\ri}{\sqrt{2}}(X^+_\alpha + X^+_{-\alpha}),
\qquad
Z^+_\alpha :=\frac{1}{\sqrt{2}}(X^+_\alpha - X^+_{-\alpha})
\quad
\forall \alpha \in {\mathfrak{R}}_+,
\nonumber\\
&& Y^-_\lambda :=\frac{\ri}{\sqrt{2}}(X^-_\lambda + X^-_{-\lambda}),
\qquad
Z^-_\lambda :=\frac{1}{\sqrt{2}}(X^-_\lambda - X^-_{-\lambda})
\quad
\forall \lambda \in {\mathfrak{W}}_+,
\label{3.15}\eea
which form an orthonormal set with respect to the scalar product $\cB$ on $\cG$.
By using the folding procedure to construct the  base elements in (\ref{3.14}),
we obtain the orthogonal decompositions
\be
\cG^+ =
\cT^+ \oplus \left( \oplus_{\alpha \in {\mathfrak{R}}_+} \bR Y^+_\alpha \right)
\oplus \left( \oplus_{\alpha \in {\mathfrak{R}}_+} \bR Z^+_\alpha \right),
\quad
\cG^- =
\cT^- \oplus \left( \oplus_{\lambda \in {\mathfrak{W}}_+} \bR Y^-_\lambda \right)
\oplus \left( \oplus_{\lambda \in {\mathfrak{W}}_+} \bR Z^-_\lambda \right).
\label{3.16}\ee

As was shown at the classical level in \cite{FP1},
certain dynamical $r$-matrices
enter in the description of the spin Sutherland type models resulting from
Hamiltonian reduction based on $\Theta$-twisted conjugations.
We recall that the
dynamical $r$-matrix associated with the involutive automorphism $\theta$
is a function $R^\theta \colon \check{\cH}^+ \rightarrow \mathrm{End}(\cA)$
defined on an open subset
$\check{\cH}^+ \subset \cH^+$.
 Its `shifted' versions,
$R^\theta_\pm  = R^\theta \pm \frac{1}{2}$,
take particularly simple form, for example,
\be
R^\theta_+(h)
= R^\theta(h) + \frac{1}{2}
= \left\{\begin{array}{ll}
\frac{1}{2} & \mbox{on}\:\cH,\\
\left(\1 - \theta\circ e^{-\ad_h}|_{{\cH}^\perp}\right)^{-1} & \mbox{on}
\:\cH^{\perp}.
\end{array}\right.
\label{3.17}\ee
Take $h := \ri q \in \check{\cT}^+$ with some $q \in \cH^+_\rr$.
Then, on the basis vectors (\ref{3.15}),
we  have
\bea
&& R^\theta(\ri q) Y^+_\alpha = \half \cot \left(\frac{\alpha(q)}{2}\right) Z^+_\alpha,
\quad
R^\theta(\ri q) Z^+_\alpha = -\half \cot \left(\frac{\alpha(q)}{2}\right) Y^+_\alpha,
\quad
\forall \alpha \in {\mathfrak{R}}_+,
\nonumber\\
&& R^\theta(\ri q) Y^-_\lambda = -\half \tan
\left(\frac{\lambda(q)}{2}\right) Z^-_\lambda, \quad R^\theta(\ri q)
Z^-_\lambda = \half \tan \left(\frac{\lambda(q)}{2}\right)
Y^-_\lambda, \quad \forall \lambda \in {\mathfrak{W}}_+.
\label{3.18}
\eea
These $r$-matrices,
which solve the classical dynamical Yang-Baxter equation on $\cH^+$ \cite{AM},
appear in the subsequent equations.

\subsection{The twisted spin Sutherland models}

Now we are ready to determine the explicit form of the objects occurring
in Proposition 2.1 for the twisted conjugation action  (\ref{3.12})  based on an
involutive diagram automorphism.

Take an arbitrary point
$e^{\ri q} \in \check{T}^\Theta$ with $q \in \cH^+_\rr$.
At this point,
the value of the infinitesimal generator of the action (\ref{3.12})
corresponding to $\xi\in \cG=T_e G$ is easily seen to be
\be
\xi^\sharp_{e^{\ri q}} =
-(\dd L_{e^{\ri q}})_e \circ (\1 - \theta\circ e^{-\ad_{\ri q}} )(\xi),
\label{3.19}\ee
where $L_g \colon y \mapsto g y$ is the left-translation on $G$
by $g \in G$, and we used that $\theta^{-1}=\theta$.
On account of (\ref{2.14}), to calculate the inertia operator we need
$\xi^\sharp$ only for $\xi \in \cK^\perp$.
In our case
\be
\cK = \cT^+,
\label{3.20}\ee
and comparison with (\ref{3.17}) shows that the linear
isomorphism (\ref{2.13}) is given by
\be
 \cK^\perp \ni \xi
\mapsto \xi^\sharp_{e^{\ri q}} =
- (\dd L_{e^{\ri q}})_e \circ R^\theta_+(\ri q)^{-1} (\xi) \in T_{e^{\ri q}} G
\qquad
\forall \ri q \in \check \cT^+.
\label{3.21}\ee
Since for the transpose of $R^\theta_+(\ri q)$ with respect to $\cB$
we have $R^\theta_+(\ri q)^T = -R^\theta_-(\ri q)$,
equation (\ref{2.14}) implies the following formula of the inertia operator:
\be
\cJ(e^{\ri q}) =
-(R^\theta_+(\ri q) R^\theta_-(\ri q))^{-1} \qquad \forall \ri q \in
\check{\cT}^+.
\label{3.22}\ee
After introducing an orthonormal basis
$\{ \ri K^-_j \}$ in $\cT^-$, the set of vectors
$\{
 \ri K^-_j, Y^+_\alpha, Z^+_\alpha, Y^-_\lambda, Z^-_\lambda\}$
is an orthonormal basis in $\cK^\perp$,
and the action of $\cJ(e^{\ri q})$ on these vectors spells as
 \bea
  & \cJ(e^{\ri q}) \ri K^-_j = 4 \ri K^-_j,
\nonumber\\
& \cJ(e^{\ri q}) Y^+_\alpha = 4 \sin^2 \left(\frac{\alpha(q)}{2}\right) Y^+_\alpha,
\quad
\cJ(e^{\ri q}) Z^+_\alpha = 4 \sin^2 \left(\frac{\alpha(q)}{2}\right) Z^+_\alpha,
\label{3.23} \\
&
\cJ(e^{\ri q}) Y^-_\lambda = 4 \cos^2 \left(\frac{\lambda(q)}{2}\right) Y^-_\lambda,
\quad
\cJ(e^{\ri q}) Z^-_\lambda = 4 \cos^2 \left(\frac{\lambda(q)}{2}\right) Z^-_\lambda.
\nonumber\eea
These relations come from (\ref{3.18}) together with the identity
$R^\theta_+ R^\theta_- = (R^\theta)^2 - \quarter$.
Since the matrix of
$\cJ(e^{\ri q})$ in the above basis is diagonal, it is easy to find
its inverse and determinant.
The non-zero entries of the matrix of
$\cJ(e^{\ri q})^{-1}$ appear in the list
\bea
& \cB(\ri K^-_k, \cJ(e^{\ri q})^{-1} \ri K^-_l) = \quarter \delta_{k, l},
\nonumber\\
& \cB(Y^+_\alpha, \cJ(e^{\ri q})^{-1} Y^+_\beta)
= \frac{1}{4 \sin^2 \left(\frac{\alpha(q)}{2}\right)} \delta_{\alpha, \beta},
\quad
\cB(Z^+_\alpha, \cJ(e^{\ri q})^{-1} Z^+_\beta)
= \frac{1}{4 \sin^2 \left(\frac{\alpha(q)}{2}\right)} \delta_{\alpha, \beta},
\label{3.24}\\
& \cB(Y^-_\lambda, \cJ(e^{\ri q})^{-1} Y^-_\mu)
= \frac{1}{4 \cos^2 \left(\frac{\lambda(q)}{2}\right)} \delta_{\lambda, \mu},
\quad
\cB(Z^-_\lambda, \cJ(e^{\ri q})^{-1} Z^-_\mu)
= \frac{1}{4 \cos^2 \left(\frac{\lambda(q)}{2}\right)} \delta_{\lambda, \mu}.
\nonumber\eea
These formulae directly give the third term in the reduced Laplace-Beltrami operator
(\ref{2.17}).

To calculate the second term in (\ref{2.17}),
we first observe from (\ref{2.16}) and the above that,
up to an irrelevant non-zero multiplicative constant
(whose sign depends on the choice of $\check \cT^+$),
the density function now has the form
\be
\delta^{\half}(e^{\ri q})
= \prod_{\alpha \in {\mathfrak{R}}_+} \sin\left(\frac{\alpha(q)}{2}\right)
\prod_{\lambda \in {\mathfrak{W}}_+} \cos\left(\frac{\lambda(q)}{2}\right).
\label{3.25}\ee
Choosing  an orthonormal basis $\{ \ri K^+_j \}$ in $\cT^+$,
we parametrize $\check{T}^\Theta$ by the diffeomorphism
\be
\check{\cT}^+ \ni \ri q = q^k \ri K^+_k \mapsto e^{\ri q} \in \check{T}^\Theta,
\label{3.26}\ee
where $\check{\cT}^+$ is an appropriate bounded open domain in $\cT^+$.
In the coordinates
$\{ q^k \}$ the Laplace-Beltrami  operator
$\Delta_{\check{T}^\Theta}$ is simply\footnote{The coordinates $q^k$
 associated with an orthonormal basis of $\cT^+$
should not be confused with the components $q_i$  used in the examples
presented at the end of this chapter and in Chapter 5.}
\be
\Delta_{\check{T}^\Theta}
= \sum_k \partial_k^2,
\label{3.27}\ee
where $\partial_k := \frac{\partial}{\partial q^k}$.
We then find that in the present case the second term of (\ref{2.17})  yields just a constant,
\be
\left(\delta^{-\half} \Delta_{\check{T}^\Theta}(\delta^\half)\right)(e^{\ri q})
= -\langle {\boldsymbol{\varrho}}_\cG^\theta, {\boldsymbol{\varrho}}_\cG^\theta \rangle
\label{3.28}\ee
with
\be
{\boldsymbol{\varrho}}_\cG^\theta
:= \half \sum_{\alpha \in {\mathfrak{R}}_+} \alpha +
 \half \sum_{\lambda \in {\mathfrak{W}}_+} \lambda,
\label{3.29}\ee
generalizing the well-known $\theta = \mathrm{id}$ case \cite{OP,Helg}.
The derivation of (\ref{3.28}) and the value of
$\langle {\boldsymbol{\varrho}}_\cG^\theta, {\boldsymbol{\varrho}}_\cG^\theta\rangle$
are elaborated in Appendix A.
The foregoing considerations are now summarized
by the following result.

\bigskip
\noindent
\textbf{Proposition 3.1.}
\emph{Consider the twisted conjugation action (\ref{3.12})
of a compact, connected, simply connected,
simple Lie group $G$ on itself,  equipped with the Riemannian metric
corresponding to the scalar product $\cB$ (\ref{3.10}) on $\cG$.
Then, using the above notations, the  reduced Laplace-Beltrami operator (\ref{2.17})
associated with a finite dimensional unitary irrep $(\rho, V_\rho)$ of $G$ takes the form
\bea
&& \Delta_{\mathrm{red}}
= \Delta_{\check{T}^\Theta} + \langle {\boldsymbol{\varrho}}_\cG^\theta, {\boldsymbol{\varrho}}_\cG^\theta \rangle
+ \quarter \sum_j \rho'(\ri K^-_j)^2
\nonumber\\
&& \qquad\quad + \quarter \sum_{\alpha \in {\mathfrak{R}}_+}
\frac{\rho'(Y^+_\alpha)^2 +\rho'(Z^+_\alpha)^2}{\sin^2 \left(\frac{\alpha(q)}{2}\right)}
+ \quarter \sum_{\lambda \in {\mathfrak{W}}_+}
\frac{\rho'(Y^-_\lambda)^2 +\rho'(Z^-_\lambda)^2}{\cos^2 \left(\frac{\lambda(q)}{2}\right)}.
\label{3.30}\eea
This operator is essentially self-adjoint on the dense domain
$\delta^\half \mathrm{Fun}(\check{T}^\Theta, V_\rho^{K})$ of the reduced
Hilbert space $L^2(\check{T}^\Theta, V_\rho^{K}, \dd \mu_{\check{T}^\Theta})$, where $K=T^\Theta$
and ${\check T}^\Theta$ is parametrized according to (\ref{3.26}).
}

\bigskip
Now we have a few remarks to make.
First, since $K=T^\Theta$ is connected, we can equivalently write $V_\rho^K$ as
$V_\rho^{\cT^+}$, which denotes the set of vectors annihilated by $\cT^+$ in the
representation $\rho'$. We shall make use of this remark in Chapter 5.
Second, as follows for example by comparison with the results in \cite{MW},
the possible domains
$\check{\cT}^+ \subset \cT^+$ that parametrize $\check T^\Theta$ in a one-to-one manner
by the exponential map
can be determined very explicitly
from the formula (\ref{3.25}). Namely, each connected component of the open set
$\{\ri q \, | \, \ri q \in \cT^+, \delta(e^{\ri q}) > 0 \} \subset \cT^+$
is an appropriate choice for $\check{\cT}^+$.
Using the coordinates $q^k$ (\ref{3.26}),  $\dd \mu_{\check{T}^\Theta} = \prod_k \dd q^k$
is the Lebesgue measure on $\check{\cT}^+$.
Third, according to Remark 2.3, one can also describe the reduced system in terms
of the larger configuration space $\hat T^\Theta$.
In this description, the wave functions are equivariant with respect to
the generalized Weyl group (\ref{2.18}), which in the present case becomes the so-called
`twisted Weyl group'
\be
W(G,T^\Theta, \Theta):= N_G(T^\Theta, I^\Theta)/T^\Theta,
\label{B.1E}\ee
where
\be
N_G(T^\Theta, I^\Theta)= \{ g\in G\,\vert\, I^\Theta_g(T^\Theta)=T^\Theta\}.
\label{B.2E}\ee
Based on \cite{W,M,MW}, we review the structure of this group in Appendix B.

We finish this section by listing the positive roots ${\mathfrak{R}}_+$ and weights
${\mathfrak{W}}_+$  and a possible choice for $\check{\cT}^+ \subset \cT^+$
for the non-trivial involutive diagram automorphisms of the classical Lie algebras.

If $\cA = D_{n+1}$, then $\cA^+ = B_n$ and the module $\cA^-$ is isomorphic
to the defining representation of $B_n$. The real Abelian subspace $\cH^+_\rr$
can be realized as real diagonal matrices of the form
\be
q = \mathrm{diag}(q_1, q_2, \ldots, q_n, 0, 0, -q_n, \ldots, -q_2, -q_1).
\label{3.31}\ee
By introducing the linear functionals $e_m \colon q \mapsto q_m$, we can
write
\be
{\mathfrak{R}}_+ = \{ e_k \pm e_l, e_m \, | \, 1 \leq k < l \leq n, 1 \leq m \leq n \},
\quad
{\mathfrak{W}}_+ = \{ e_m \, | \, 1 \leq m \leq n \}.
\label{3.32}\ee
A possible choice for $\check{\cT}^+ \subset \cT^+$ is given by the set
\be
\check{\cT}^+ = \{ \ri q \, | \, 0 < q_n < q_{n-1} < \cdots <q_2 < q_1 <  \pi \}.
\label{3.33}\ee

If $\cA = A_{2n - 1}$, then $\cA^+ = C_n$ and $\cH^+_\rr$ can be
realized as diagonal matrices of the form
\be
q = \mathrm{diag}(q_1, q_2, \ldots, q_n, -q_n, \ldots, -q_2, -q_1).
\label{3.34}\ee
Using the functionals $e_m \colon q \mapsto q_m$, we can write
\be
{\mathfrak{R}}_+ = \{ e_k \pm e_l, 2 e_m \, | \, 1 \leq k < l \leq n, 1 \leq m \leq n \},
\quad
{\mathfrak{W}}_+ = \{ e_k \pm e_l \, | \, 1 \leq k < l \leq n \}.
\label{3.35}\ee
For $\check{\cT}^+$ we can choose the set
\be
\check{\cT}^+ = \{ \ri q \, | \, 0 < q_n < q_{n-1} < \cdots <q_2 < q_1 <  \pi,\,\,
0 < q_1 + q_2 <  \pi \}.
\label{3.36}\ee

For $\cA = A_{2n}$ one has $\cA^+ = B_n$. Now $\cH^+_\rr$ can be realized as
\be
q = \mathrm{diag}(q_1, q_2, \ldots, q_n, 0, -q_n, \ldots, -q_2, -q_1),
\label{3.37}\ee
and with $e_m \colon q \mapsto q_m$ we have
\bea
&& {\mathfrak{R}}_+ = \{ e_k \pm e_l, e_m \, | \, 1 \leq k < l \leq n, 1 \leq m \leq n \},
\nonumber\\
&& {\mathfrak{W}}_+ = \{ e_k \pm e_l, e_m, 2 e_m \, | \, 1 \leq k < l \leq n, 1 \leq m \leq n \}.
\label{3.38}\eea
A possible choice for $\check{\cT}^+$ is the bounded open domain
\be
\check{\cT}^+ = \{ \ri q \, | \, 0 < q_n < q_{n-1} < \cdots <q_2 < q_1 <  \frac{\pi}{2} \}.
\label{3.39}\ee

\section{How to diagonalize the  reduced Hamiltonians?}
\setcounter{equation}{0}

We below explain how the diagonalization of the `twisted spin Sutherland Hamiltonians'
$\Delta_{\mathrm{red}}$ given by Proposition 3.1
can be performed in principle in terms of certain Clebsch-Gordan problems
in the representation theory of the underlying symmetry group $G$.

Let us start by noting that
the Hilbert space $L^2(G, \dd\mu_G)$ carries the unitary representation
$U_\theta(g)$ of the group $G$ defined by
\be
U_\theta(g) \psi := \psi\circ I_{g^{-1}}^\Theta
\label{4.1}\ee
for any `wave function' $\psi$ and $g\in G$.
Let $L^+$ be
the set of dominant integral weights of $G$ and for
any $\Lambda\in L^+$ denote
by $(\rho_\Lambda, V_\Lambda)$
the corresponding irreducible unitary highest weight
representation\footnote{We should
have written  $V_{\rho_\Lambda}$ according to our general notation used before,
but here we simplify this to $V_\Lambda$.},
of dimension $d_\Lambda$.
By arbitrarily fixing a  basis in $V_\Lambda$,  introduce the matrix elements
$\rho_\Lambda^{a,b} \in L^2(G, \dd\mu_G)$  spanning the subspace
\be
F^\Lambda := \operatorname{span}\{ \rho_\Lambda^{a,b} \,\vert\, a,b=1,\ldots, d_\Lambda\,\}
\label{4.2}\ee
of $L^2(G, \dd\mu_G)$.
By the Peter-Weyl theorem, one has the orthogonal direct sum decomposition
\be
L^2(G,\dd\mu_G) = \oplus_{\Lambda \in L^+} F^\Lambda.
\label{4.3}\ee
The restriction of the  representation $U_\theta$ to $F^\Lambda$
can be written as the following tensor product of irreps:
\be
F^\Lambda = V_{(\Lambda \circ \theta^{-1})^*} \otimes V_\Lambda,
 \label{4.4}\ee
where $\Lambda^*$ denotes the highest weight
of the complex conjugate of the representation $\rho_\Lambda$ and,
as $\theta$ gives  a Cartan preserving automorphism of the
complexification of $\cG$, $\Lambda \circ \theta^{-1} \in L^+$ is a well-defined
functional on the Cartan subalgebra.
With these notations, using that $\theta$ is an involution, we have
\be
L^2(G,\dd\mu_G) = \oplus_{\Lambda \in L^+}  V_{(\Lambda \circ \theta)^*} \otimes V_\Lambda.
\label{4.5}\ee

Focusing now on our  Hamiltonian reduction problem,  let us set
\be
(\rho,V_\rho):= (\rho_\nu,V_\nu)
\label{4.6}\ee
with some highest weight $\nu$ (see footnote 3).
Then the reduced Hilbert space takes the form
\be
L^2(G, V_\nu, \dd\mu_G)^G = \oplus_{\Lambda\in L^+}
(V_{(\Lambda \circ \theta)^*} \otimes V_\Lambda \otimes V_\nu)^G.
\label{4.7}\ee
Therefore we have to find the singlets, i.e., the $G$-invariant states,
 in the threefold tensor products in (\ref{4.7}).
Any such singlet state,  say
\be
\cF_{\Lambda,\nu} \in (V_{(\Lambda \circ \theta)^*} \otimes V_\Lambda \otimes V_\nu)^G,
\label{4.8}\ee
is in effect a $V_\nu$-valued $I^\Theta$-equivariant function on $G$.

The point is that \emph{the functions $\cF_{\Lambda,\nu}$ are
eigenstates of the Laplace-Beltrami operator $\Delta_G$},
\be
\Delta_G \cF_{\Lambda,\nu}  =
- \langle \Lambda + 2 {\boldsymbol{\varrho}}_\cG, \Lambda\rangle \cF_{\Lambda,\nu},
\label{4.9}\ee
since $\Delta_G$ corresponds to the quadratic Casimir operator of the Lie algebra $\cG$,
that survive the reduction.
${\boldsymbol{\varrho}}_\cG$ stands for the sum of the fundamental weights
of the complexification, $\cA$, of $\cG$.
More generally,  take any left $G$-invariant, formally self-adjoint scalar
differential operator over $G$ induced by a corresponding element of
the center of the universal enveloping algebra of $\cA$.
Any such differential operator,  say $\hat P$, takes a constant
value on $F^\Lambda$,
\be
{\hat P} \cF= p(\Lambda+{\boldsymbol{\varrho}}_\cG) \cF \qquad \forall \cF\in F^\Lambda,
\label{4.10}\ee
with an associated Weyl invariant polynomial function $p$ on the dual of the Cartan subalgebra.
The map sending $\hat P$ to $p$ is the so-called Harish-Chandra isomorphism \cite{Knapp}.
(A rather explicit formula for the eigenvalue $p(\Lambda +{\boldsymbol{\varrho}}_\cG)$
of the Casimir operator
$\hat P$ is contained in \cite{Berezin}.)
It is standard to show that
these commuting differential operators
induce commuting self-adjoint \lq Hamiltonians' for the reduced quantum system.

Returning  to (\ref{4.7}),
 one has the Clebsch-Gordan series
\be
V_\Lambda \otimes V_\nu = \oplus_{\lambda \in L^+} N_{\Lambda,\nu}^\lambda V_\lambda,
\label{4.11}\ee
where $N_{\Lambda,\nu}^\lambda\in \bZ_+$ are often called Littlewood-Richardson numbers.
In principle, they can be found algorithmically  since the irreducible characters
are known.
In order to find the spectra of the reduced Laplace-Beltrami operator (as well as the spectra
of the higher Casimirs) on $L^2(G,V_\nu, \dd\mu_G)^G$, we need the numbers
 \be
 \operatorname{dim} (V_{(\Lambda \circ \theta)^*} \otimes V_\Lambda \otimes V_\nu)^G =
 N_{\Lambda,\nu}^{\Lambda\circ \theta}.
\label{4.12}\ee
In fact, it follows from the definition of $\Delta_{\mathrm{red}}$ that
\be
\operatorname{spectrum}\left(\Delta_{\mathrm{red}}\right)
=\{ - \langle \Lambda + 2{\boldsymbol{\varrho}}_\cG, \Lambda \rangle\,\vert\, \Lambda \in L^+,
\,
N_{\Lambda, \nu}^{\Lambda \circ \theta}\neq 0\}.
\label{4.13}\ee
The Littlewood-Richardson numbers  determine
 the multiplicities of the eigenvalues of $\Delta_{\mathrm{red}}$ as well.
It is an archetypical Clebsch-Gordan problem to find  \emph{explicitly}
the pairs $(\Lambda,\nu)$
for which  $N_{\Lambda,\nu}^{\Lambda\circ \theta}\neq 0$.
It is even more difficult to obtain the explicit form of the
states $\cF_{\Lambda,\nu}$, since they involve
the Clebsch-Gordan coefficients themselves.
The diagonalization of the reduced Hamiltonian
boils down to these finite dimensional, purely algebraic problems.

The eigenfunctions $\cF_{\Lambda,\nu}$
`descend' to certain trigonometric polynomials through the reduction procedure.
To see this, let us take a basis $v^\Lambda_a$ in $V_\Lambda$ consisting of  weight vectors
with respect to $\cH^+$, with weight $\mu_a$:
\be
\rho'_\Lambda(\ri q) v_a^\Lambda = \ri \mu_a(q) v_a^\Lambda
\qquad
\forall q \in \cH^+_\rr.
\label{4.14}\ee
Similarly, take a basis $u_k^\nu$ in $V_\nu^{T^\Theta}$.
The restricted eigenfunction of $\Delta_G$,
\be
{\hat f}_{\Lambda,\nu}:= \cF_{\Lambda,\nu}\vert_{T^\Theta},
\label{4.15}\ee
is a linear combination of functions of the form
\be
(v_a^\Lambda , \rho_\Lambda(e^{\ri q}) v_b^\Lambda) u_k^\nu,
\label{4.16}\ee
where $\mu_a + \mu_b =0$.
Therefore we can write
\be
{\hat f}_{\Lambda,\nu}(e^{\ri q})= \sum_{a,k} C_{a,k} e^{\ri \mu_a(q)} u^\nu_k
\label{4.17}\ee
with some constants $C_{a,k}$.
The sum runs over all the $\cH^+$-weights of the representation space $V_\Lambda$
and a basis of $V_\nu^{T^\Theta}$.
It is also worth noting that,
as a result of the $G$-equivariance of $\cF_{\Lambda,\nu}$,
the  $\hat f_{\Lambda,\nu}$ are  Weyl-equivariant functions on $T^\Theta$, where
the relevant generalized Weyl group is $W(G,T^\Theta,\Theta)$ (\ref{B.1E}),
which acts naturally both on $V_\nu^{T^\Theta}$ and on $T^\Theta$.
Of course,  the eigenfunctions of the spin Sutherland operator $\Delta_{\mathrm{red}}$, obtained
as a similarity transform of the restriction of $\Delta_G$,  also include the
pre-factor  $\delta^{\frac{1}{2}}$ (\ref{3.25})
in front of the above function $\hat f_{\Lambda,\nu}$,
or rather in front of
$f_{\Lambda,\nu}:=  \hat f_{\Lambda,\nu}\vert {\check T^\Theta}$.
See also the remarks after
Proposition 2.1 and Appendix B.

\section{Some examples with explicitly computable spectra}
\setcounter{equation}{0}

In a few distinguished cases the Littlewood-Richardson numbers (\ref{4.12})
can be determined explicitly due to `standard plethysm'
rules in representation theory \cite{FH}.
For $\cG=su(N)$, such are the cases
\be
\nu = k \lambda_1 \quad\hbox{or}\quad k \lambda_{N-1},
\quad \hbox{and}\quad
\nu = \lambda_a
\quad\hbox{for}\quad
a=2,\ldots, (N-2),
\label{5.1}\ee
where from now on we denote by $\lambda_a$,  $a=1,\ldots, \operatorname{rank}(\cG)$, the
fundamental weights.
Indeed, in these cases one has the so-called Pieri formulae for the decomposition
of the tensor products $V_\Lambda \otimes V_\nu$.
The first two cases are essentially equivalent, being related by the Dynkin diagram
symmetry of $A_{N-1}$.
Below we concentrate on the first case, since it contains also an arbitrary parameter $k\in \bZ_+$.
Throughout this chapter, we use
$ {\boldsymbol{\varrho}}_{su(N)}= \sum_{a=1}^{N-1} \lambda_a$,
which is the $\theta=\mathrm{id}$ special case of ${\boldsymbol{\varrho}}_{su(N)}^\theta$ in ({\ref{3.29}).
We let  $T_N\subset SU(N)$, $\cT_N \subset su(N)$ stand for the standard maximal torus
and its Lie algebra, and denote the root system of $sl(N,\bC)$ by $\Phi^N$.
We fix the invariant bilinear form on $sl(N,\bC)$ to be
\be
\langle X,Y \rangle := \operatorname{tr}(XY)
\qquad
\forall X,Y \in sl(N,\bC).
\label{5.2}\ee

\subsection{Oscillator realization of  $V_{k\lambda_1}$  for $su(N)$ and corresponding models}

In this subsection we recall the oscillator realization of $V_{k\lambda_1}$ and
spell out the corresponding models both for the trivial and for the non-trivial
Dynkin diagram automorphisms of $su(N)$.

Let us introduce $N$ pairs of bosonic annihilation and creation operators, $a_i$ and $a_i^\dagger$,
\be
[a_i, a_j^\dagger] = \delta_{i, j}
\qquad
(1 \leq i, j \leq N).
\label{5.3}\ee
These operators act on the Fock space, $\cV$, spanned by the basis
\be
\prod_{i=1}^N (a_i^\dagger)^{n_i} v_0
\qquad (\forall  n_i \in \bZ_+),
\label{5.4}\ee
where $v_0$ is the vacuum vector annihilated by the $a_i$.
We also need the `number operators'
\be
\hat n_i := a_i^\dagger a_i,
\label{5.5}\ee
whose eigenvalues are the $n_i$ in (\ref{5.4}).
The Lie algebra $gl(N,\bC)$ is represented on $\cV$ by
\be
E_{i,j}\mapsto \rho'(E_{i,j}):= a_i^\dagger a_j.
\label{5.6}\ee
One has
$\cV= \oplus_{k\in \bZ_+} \cV_k$,
where $\cV_k$ is the eigensubspace of
the `total number operator' $\left(\sum_{i=1}^N \hat n_i\right)$
with eigenvalue $k$, which is an invariant subspace for $gl(N,\bC)$.
Restriction of $\rho'$ to $\cV_k$, and to the subalgebra $su(N)\subset gl(N,\bC)$, yields
a model of the highest weight representation $V_{k\lambda_1}$.
We display this fact symbolically as
\be
V_{k\lambda_1} \cong \cV_k.
\label{5.7}\ee

It is easy to see from (\ref{5.6}) that $\cV_k^{\cT_N}$ is non-zero if and only if
$k=\gamma N$ for some
$\gamma\in \bZ_+$ and
\be
\cV_{\gamma N}^{\cT_N} = \bC \prod_{i=1}^N (a_i^\dagger)^\gamma v_0
\label{5.8}\ee
is one-dimensional. It also follows from (\ref{5.6})  that
\be
\left(\rho'(E_{i,j}) \rho'(E_{j,i})\right)
\vert_{\cV_{\gamma N}^{\cT_N}}=
\gamma(\gamma+1)\operatorname{id}_{\cV_{\gamma N}^{\cT_N}}
\qquad
\forall i\neq j.
\label{5.9}\ee
Thus in the non-twisted case (with
$\cG=su(N)$, $\theta=\mathrm{id}$ and $V=V_{k\lambda_1}$),  $\Delta_{\mathrm{red}}$ in
(\ref{3.30}) yields
\be
\Delta_{\mathrm{red}}^\gamma=
\Delta_{\check T_N}
+\langle {\boldsymbol{\varrho}}_{su(N)}, {\boldsymbol{\varrho}}_{su(N)} \rangle - \frac{1}{2}
\sum_{\alpha \in\Phi^N_+}\frac{\gamma(\gamma+1)}
{\sin^2\alpha(\frac{q}{2})},
\label{5.10}\ee
which is the standard Sutherland operator with coupling constant determined by $\gamma\in \bZ_+$
($\check T_N$ is parametrized by $e^{\ri q}$ with $q$ in a Weyl alcove).
This result is of course  well-known \cite{EFK,Poly}.

Turning to the twisted spin Sutherland models,
first note that the non-trivial automorphism
of $su(N)$ can be defined on the generators $E_{i,j}\in gl(N,\bC)$ as
\be
\theta(E_{i, j}) := -(-1)^{i + j} E_{N + 1 - j, N + 1 - i}.
\label{5.11}\ee
Combining this with (\ref{5.6}), we obtain that if $N=2n$ is even, then
\be
\cV_{k}^{\cT_{2n}^+}=\operatorname{span}\left\{
\prod_{i=1}^n (a_i^\dagger a_{2n+1-i}^\dagger)^{n_i} v_0 \,\Big|\, 2\sum_{i=1}^n n_i = k\,
\right\}.
\label{5.12}\ee
The non-triviality of this space requires $k$ to be even, and
\be
\operatorname{dim}( \cV_{2 \kappa}^{\cT_{2n}^+} ) =
{{\kappa + n-1}\choose{n-1}},
\qquad
\forall \kappa, n\in \bZ_+,\, n\geq 2.
\label{5.13}\ee

If $N=(2n+1)$ is odd, then the index $i=(n+1)$ is special and we introduce the notation
\be
c:= a_{n+1},\quad  c^\dagger = a_{n+1}^\dagger,
\quad
\hat m:= c^\dagger c.
\label{5.14}\ee
In this case we find that
\be
\cV_{k}^{\cT_{2n+1}^+}=\operatorname{span}\left\{
(c^\dagger)^{m} \prod_{i=1}^n (a_i^\dagger a_{2n+2-i}^\dagger)^{n_i} v_0 \,\Big|\,
m + 2\sum_{i=1}^n n_i = k\, \right\},
\label{5.15}\ee
and
\be
\operatorname{dim}( \cV_{k}^{\cT_{2n+1}^+}) =
\sum_{\kappa =0}^{[k/2]}
{{\kappa + n-1}\choose{n-1}},
\qquad
\forall k, n\in \bZ_+,\, n\geq 1.
\label{5.16}\ee
This dimension formula was obtained by  setting
$2\kappa := 2\sum_{i=1}^n n_i = k - m$.

Based on (\ref{5.6}),
it is a matter of straightforward calculation to derive from Proposition 3.1
explicit  formulae for the reduced Laplace-Beltrami operators in the above
 cases. We give a brief outline of this calculation in Appendix C, and here only
display the result.
For this purpose, for any $N=2n$ or $N=(2n+1)$, we define the operators
\bea
&&\cA_{i, j} := 2 \hat{n}_i \hat{n}_j + \hat n_i + \hat n_j,
\qquad 1\leq i\leq j\leq n,
\label{5.17}\\
&&\cB_{i, j} := (-1)^{i + j}(a_i^\dagger a_{N+1-i}^\dagger a_j a_{N+1-j}
+ a_i a_{N+1-i} a_j^\dagger a_{N+1-j}^\dagger),
\qquad  1\leq i<j\leq n.
\nonumber\eea
If $N=(2n+1)$, then using (\ref{5.14}) we also define
\bea
&& \cC_i
:= 2 \hat{n}_i \hat m  +\hat{m} + \hat{n}_i,
\qquad  1\leq i\leq n,
\label{5.18}\\
&& \cD_i
:= (-1)^{i + n}(a_i^\dagger a_{2n+2-i}^\dagger c c
+ a_i a_{2n+2-i} c^\dagger c^\dagger),
\qquad  1\leq i\leq n.
\nonumber\eea
These operators act on the space $\cV_k^{\cT_N^+}$, where
the reduced wave functions take their values.

\bigskip
\noindent
{\bf Proposition 5.1.} \emph{Choosing $N=2n$ ($n\geq 2$) even and the  representation
$V_{2\kappa \lambda_1}\cong \cV_{2\kappa}$ ($\kappa \in \bZ_+$) of $G=SU(N)$, the reduced
Laplace-Beltrami operator (\ref{3.30})  corresponding to the diagram automorphism (\ref{5.11})
is given by the explicit formula
\bea
&& \Delta_{\mathrm{red}}
=
\frac{1}{2}\sum_{i=1}^n \frac{\partial^2}{\partial q_i^2}
+ \langle {\boldsymbol{\varrho}}^\theta_{su(2n)}, {\boldsymbol{\varrho}}^\theta_{su(2n)} \rangle
+\frac{\kappa^2}{2n}
- \frac{1}{2} \sum_{i = 1}^n \hat{n}_i^2
-\frac{1}{4} \sum_{i=1}^n \frac{\cA_{i,i}}{\sin^2 \left( q_i \right)}
\label{5.19} \\
&&  \quad -\frac{1}{4} \sum_{1\leq i < j\leq n} \left[\frac{\cA_{i,j} -
\cB_{i,j}}{\sin^2 \left( \frac{q_i - q_j}{2} \right)}
+ \frac{\cA_{i,j} + \cB_{i,j}}{\cos^2 \left( \frac{q_i - q_j}{2} \right)}
+ \frac{\cA_{i,j} + \cB_{i,j}}{\sin^2 \left( \frac{q_i + q_j}{2} \right)}
+ \frac{\cA_{i,j} - \cB_{i,j}}{\cos^2 \left( \frac{q_i + q_j}{2} \right)}\right],
\nonumber
\eea
where we use (\ref{5.17}).
The constant
$\langle {\boldsymbol{\varrho}}^\theta_{su(2n)}, {\boldsymbol{\varrho}}^\theta_{su(2n)} \rangle$
is given by (\ref{A.7}) in  Appendix A,  and  the coordinates $q_i$ can be taken to
vary in the domain (\ref{3.36}).
}

\bigskip
\noindent
{\bf Proposition 5.2.}
\emph{Choosing $N=(2n+1)$ ($n\geq 1$) odd and the  representation
$V_{k\lambda_1}\cong \cV_{k}$ ($k \in \bZ_+$) of $G=SU(N)$, the reduced
Laplace-Beltrami operator (\ref{3.30})  corresponding to the diagram automorphism (\ref{5.11})
is given by the explicit formula
\bea
&& \Delta_{\mathrm{red}}
=\frac{1}{2}\sum_{i=1}^n \frac{\partial^2}{\partial q_i^2}
+
\langle {\boldsymbol{\varrho}}^\theta_{su(2n+1)}, {\boldsymbol{\varrho}}^\theta_{su(2n+1)} \rangle
- \frac{1}{2}
\sum_{i = 1}^n (\hat{n}_i - \hat{m})^2
\label{5.20}\\
&& \quad
+\frac{1}{2n} \left[ \sum_{i = 1}^n ( \hat{n}_i -  \hat{m}) \right]^2  -\frac{1}{4}
\sum_{i=1}^n \left[  \frac{\cC_i + \cD_i}{\sin^2\left(\frac{q_i}{2}\right)}
+\frac{\cC_i - \cD_i}{\cos^2\left(\frac{q_i}{2}\right)}
+\frac{\cA_{i, i}}{\cos^2\left(q_i\right)}\right]
\nonumber\\
&&  \quad -\frac{1}{4} \sum_{1\leq i < j\leq n}
\left[\frac{\cA_{i,j} -\cB_{i,j}}{\sin^2 \left( \frac{q_i - q_j}{2} \right)}
+ \frac{\cA_{i,j} + \cB_{i,j}}{\cos^2 \left( \frac{q_i - q_j}{2} \right)}
+ \frac{\cA_{i,j} - \cB_{i,j}}{\sin^2 \left( \frac{q_i + q_j}{2} \right)}
+ \frac{\cA_{i,j} + \cB_{i,j}}{\cos^2 \left( \frac{q_i + q_j}{2} \right)}\right],
\nonumber
\eea
where we use the operators defined in (\ref{5.17}), (\ref{5.18}).
The constant
$\langle {\boldsymbol{\varrho}}^\theta_{su(2n+1)}, {\boldsymbol{\varrho}}^\theta_{su(2n+1)} \rangle$
 is evaluated in (\ref{A.7}) and now the coordinates $q_i$ can be taken to
 vary in the domain (\ref{3.39}).
}

\bigskip
The above formulae of $\Delta_{\mathrm{red}}$  can be somewhat simplified  by
using obvious trigonometry, like $4 \sin^2 \frac{x}{2} \cos^2 \frac{x}{2} = \sin^2 x$.
The operators $\cA_{i,j}$, $\cB_{i,j}$, $\cC_i$ and $\cD_i$ act on the respective spaces
$\cV_{k}^{\cT^+_N}$, where  $\cA_{i,j}$ and $\cC_i$ are diagonal, whilst  $\cB_{i,j}$ and
$\cD_i$ are off-diagonal in the natural bases that appear in
equations (\ref{5.12}) and (\ref{5.15}).
Subsequently,  we  shall determine the spectra of the self-adjoint matrix differential operators
$\Delta_{\mathrm{red}}$ given by (\ref{5.19}) and (\ref{5.20}).

\subsection{Determination of the spectra from the  Pieri formula}

For $\cG=su(N)$, write an arbitrary $\Lambda\in L^+$ in the form
\be
\Lambda =\sum_{i=1}^{N-1} M^i \lambda_i
\qquad
M^i \in \bZ_+.
\label{5.21}\ee
Then the so-called \emph{Pieri formula} (see e.g.~\cite{FH})  says that precisely those
\be
\lambda = \sum_{i=1}^{N-1} m^i \lambda_i \qquad m^i \in \bZ_+
\label{5.22}\ee
appear in the tensor product $V_\Lambda \otimes V_{k\lambda_1}$ for which
\be
m^i = M^i + C^i - C^{i+1}
\quad
\forall i=1,\ldots, (N-1)
\label{5.23}\ee
with some
$(C^1,C^2,\ldots, C^{N-1}, C^N)\in \bZ_+^N$
such that
\be
C^{i+1} \leq M^i \quad \forall i=1,\ldots, (N-1),
\qquad
\sum_{i=1}^N C^i =k.
\label{5.24}\ee
Note that $N^\lambda_{\Lambda, k\lambda_1}=1$ for all $\lambda$ for which
$N^\lambda_{\Lambda, k\lambda_1}\neq 0$.
The spectra of the reduced Laplace-Beltrami operators displayed in subsection 5.1
are easily found by utilizing  the Pieri formula.

\subsubsection{The standard Sutherland model with integer couplings}

In the non-twisted $su(N)$ case, for $\theta=\operatorname{id}$  and
$\nu =k \lambda_1$,
the determination of the spectrum (\ref{4.13})
boils down to finding all $\Lambda\in L^+$ for which
\be
N^\Lambda_{\Lambda, k\lambda_1}=1.
\label{5.25}\ee
From the Pieri formula (5.23) with $m^i=M^i$ we immediately
get that
\be
C^1=C^2=\ldots =C^N := \gamma \in \bZ_+
\label{5.26}\ee
and therefore
$k= \gamma N$
with some
$\gamma\in \bZ_+$, which also follows from the condition
$\operatorname{dim}(V_{k\lambda_1}^{\cT_N})\neq 0$ as we saw.
The first part of (\ref{5.24}) then says that we must have
\be
M^i = \gamma + \mu^i
\quad\hbox{with some arbitrary}\quad \mu^i\in \bZ_+.
\label{5.27}\ee
This means that \emph{the admissible weights $\Lambda$ are precisely the weights  of the form}
\be
\Lambda = \gamma {\boldsymbol{\varrho}}_{su(N)} + \mu,
\quad \forall \mu=\sum_{i=1}^{N-1} \mu^i \lambda_i \in L^+.
\label{5.28}\ee
We have already seen  that
$\operatorname{dim}(V_{\gamma N \lambda_1}^{\cT_N})=1$ by (\ref{5.8}),
and have derived the formula (\ref{5.10}) of  $\Delta_{\mathrm{red}}$
that arises in this case.
We now rewrite this scalar Schr\"odinger operator in the form
\be
\Delta_{\mathrm{red}}^{\gamma}=
\frac{1}{4}\sum_{i=1}^{N-1}
\left(\frac{\partial}{\partial x^i}\right)^2
+\langle {\boldsymbol{\varrho}}_{su(N)}, {\boldsymbol{\varrho}}_{su(N)} \rangle - \frac{1}{2}
\sum_{{1\leq i < j\leq N}}\frac{\gamma(\gamma+1)}
{\sin^2(x_i-x_j)},
\label{5.29}\ee
where  $x:=  \frac{q}{2}$ with the real Cartan variable $q\in \cH_\rr$ introduced before,
and the $x^{i}$ denote the components of $x=\mathrm{diag}(x_1,\ldots, x_N)$
in an orthonormal basis of  $\cH_\rr$
(here $\cT_N=\ri \cH_\rr \subset su(N)$, see also footnote 2).

Recall that
the standard $N$-particle Sutherland Hamiltonian, after separation of the
center of mass and introducing dimensionless variables,
can be written as follows:
\be
H_{\mathrm{standard}}^{N,g} = -\frac{1}{2}\sum_{i=1}^{N-1}
\left(\frac{\partial}{\partial x^{i}}\right)^2 + \sum_{{1\leq i < j\leq N}}
\frac{g(g-1)}{\sin^2(x_i-x_j)},
\label{5.30}\ee
where  the domain of $x$ is bounded by the singular locus of the potential.
By setting $g:= (\gamma+1)$,
comparison shows that
\be
H_{\mathrm{standard}}^{N,g} = - 2\Delta_{\mathrm{red}}^{\gamma}  +
2 \langle {\boldsymbol{\varrho}}_{su(N)}, {\boldsymbol{\varrho}}_{su(N)}\rangle.
\label{5.31}\ee
It follows from (\ref{5.28})  that
\be
\operatorname{spectrum}(\Delta_{\mathrm{red}}^{\gamma})= \{ -
\langle (\gamma{\boldsymbol{\varrho}}_{su(N)} + \mu) + 2{\boldsymbol{\varrho}}_{su(N)},
(\gamma{\boldsymbol{\varrho}}_{su(N)} + \mu)\rangle\,\vert\,
\mu \in L^+\,\}.
\label{5.32}\ee
Combining the last two equations we  recover the well-known formula (see e.g.~\cite{Sas})
\be
\operatorname{spectrum}(H_{\mathrm{standard}}^{N,g}) =
\{ 2 \langle \mu + g{\boldsymbol{\varrho}}_{su(N)}, \mu +
g{\boldsymbol{\varrho}}_{su(N)} \rangle\,\vert\,
\mu \in L^+\,\}
\label{5.33}\ee
as a consequence of representation theory \cite{EFK}; for integer coupling constants $g>0$.

\subsubsection{The spectrum in the twisted $su(N)$ case with $\nu=k\lambda_1$}

The non-trivial diagram automorphism $\theta$ satisfies
\be
\Lambda^* = \Lambda\circ \theta
\label{5.34}\ee
for any dominant integral weight of $su(N)$.
This means that
\be
\Lambda^* = \sum_{i=1}^{N-1} M^{N-i} \lambda_i
\qquad\hbox{for}\qquad
\Lambda = \sum_{i=1}^{N-1} M^i \lambda_i,
\quad M^i\in \bZ_+.
\label{5.35}\ee
Therefore, to find the spectrum (\ref{4.13}) for the operators
$\Delta_{\mathrm{red}}$ in (\ref{5.19}) and (\ref{5.20}), we have to determine
the admissible weights  $\Lambda$ for which
\be
N^{\Lambda^*}_{\Lambda, k\lambda_1}=1.
\label{5.36}\ee

\bigskip
\noindent
\textbf{Theorem 5.3.} \emph{The dominant integral weights $\Lambda$ of $su(N)$ that
satisfy (\ref{5.36}) have the form
\be
\Lambda = \mu +\chi,
\label{5.37}\ee
where $\mu$ is an arbitrary self-conjugate dominant integral weight, $\mu^*=\mu$,
and $\chi$ reads
\be
\chi:= \sum_{i=1}^{N-1} C^{i+1} \lambda_i
\label{5.38}\ee
with some $(C^1,\ldots, C^{N}) \in \bZ_+^N$ subject to the requirements
\be
C^{N+1-i} = C^i
\quad
\forall i=1,\ldots, N
\quad\hbox{and}\quad
\sum_{i=1}^N C^i = k.
\label{5.39}\ee
The number of distinct solutions for $\chi$ equals $\operatorname{dim}(V_{k\lambda_1}^{\cT^+_N})$,
which in particular means that $k$ must be even if $N\geq 4$ is even.
The spectrum of the corresponding twisted spin Sutherland Hamiltonian,
given by (\ref{5.19}) or (\ref{5.20}), is provided by
(\ref{4.13}) with these weights $\Lambda$.}

\bigskip
\noindent
\textbf{Proof.}
Taking any $\Lambda= \sum_{i=1}^{N-1} M^i \lambda_i$ satisfying (\ref{5.36}),
the Pieri formula (\ref{5.23}) implies that
\be
M^{N-i} = M^i + C^i - C^{i+1}
\label{5.40}\ee
for some $C^i\in \bZ_+$ subject to (\ref{5.24}).
If we define
\be
\mu^i:= M^i - C^{i+1},\qquad i=1,\ldots, N-1,
\label{5.41}\ee
then we see from (\ref{5.24}) that $\mu^i \in \bZ_+$, and (\ref{5.40}) gives
\be
\mu^i = \mu^{N-i} + (C^{N+1-i} - C^i),
\qquad
\forall i=1,\ldots, N-1.
\label{5.42}\ee
Replacing $i$ by $(N-i)$, we obtain
\be
\mu^i=\mu^{N-i} + (C^{(N+1)-(i+1)} - C^{i+1}), \quad \forall i=1,\ldots, (N-1).
\label{5.43}\ee
If we now define
\be
\vartheta^i:= C^{N+1-i} - C^i \quad \forall i =1,\ldots, N,
\label{5.44}\ee
then we can read off from (\ref{5.42}) and (\ref{5.43}) that $\vartheta^i = \vartheta^{i+1}$,
and hence
\be
\vartheta^1 = \vartheta^2 = \ldots =\vartheta^N.
\label{5.45}\ee
But by its very definition $\vartheta^i$ satisfies
\be
\vartheta^{N+1-i}= C^{i}- C^{N+1-i}= - \vartheta^i,
\label{5.46}\ee
and therefore $\vartheta^i =0$, that is (\ref{5.39}) holds.
By substituting this back into equation (\ref{5.43}), we obtain
$\mu^{i} = \mu^{N-i}$.
Thus (\ref{5.41}) yields the required formula (\ref{5.37}) with
\be
\mu:= \sum_{i=1}^{N-1} \mu^i\lambda_i,
\quad
\chi:= \sum_{i=1}^{N-1} C^{i+1} \lambda_i,
\label{5.47}\ee
where $\mu^*=\mu$ and $\chi$ satisfies (\ref{5.39}).
We have just shown that all `admissible weights' must have the form  of $\Lambda$ in
(\ref{5.37}), and it is trivial to check that any
$\Lambda$ of this form satisfies (\ref{5.36}) indeed.
The claim about the number of  solutions for $\chi$  follows by obvious
comparison of (\ref{5.39}) with the bases of $\cV_{k\lambda_1}^{\cT_N^+}$ furnished by
(\ref{5.12}) and (\ref{5.15}).
\emph{Q.E.D.}

\section{Conclusion}
\setcounter{equation}{0}

In this paper we first reviewed the quantum Hamiltonian symmetry reductions of the free
particle on a Riemannian manifold  governed by the Laplace-Beltrami operator.
We assumed that the underlying symmetry group $G$
is compact and it acts in polar manner in the sense of Palais and Terng \cite{PT}.
This strong assumption permits to derive nice results in a unified way and
it holds in many important examples.
For instance, it holds for
the so-called Hermann actions (\ref{3.2}), which are hyperpolar and lead to spin Sutherland type
models in general.

In the main text we focused on the group action
defined by twisted conjugations (\ref{3.12}), and described the
corresponding twisted spin Sutherland models in Proposition 3.1.
Then we explained how to diagonalize these reduced Hamiltonians in principle, and
have characterized their spectra explicitly in Theorem 5.3
for certain special cases associated with $G=SU(N)$.
These special cases were obtained by using the symmetric tensorial
powers of the defining representation of $SU(N)$
in the definition of the reduction, and
we also presented  `oscillator realizations' of the
resulting  Hamiltonians  in Propositions 5.1 and 5.2.

The classical analogues of the twisted spin Sutherland models have  been
investigated in \cite{FP1}.
Further related studies of classical Hamiltonian reduction can be found in
\cite{Hoch,FP2} and references therein.
A comparison  between
the outcomes of the classical and quantum Hamiltonian reductions under polar actions
in general was given in \cite{FP_polar}.

The reduced systems that feature in Proposition 2.1
possess a hidden generalized Weyl group symmetry, as was pointed out in Remark 2.3.
The structure of the generalized Weyl group belonging to the twisted conjugations
(\ref{3.12}) has been clarified in \cite{M,MW}, and we outline this in Appendix B
giving an explicit description in the $SU(N)$ cases.
The other two appendices contain some technical details relegated from the main text.

There remain several open problems requiring future work.
First, it would be interesting to find the eigenfunctions for the models
studied in Chapter 5.
We note that
generalized spin Sutherland models with similar interaction potentials,
but different spin variables, have been constructed by Finkel \emph{et al}
\cite{Finkel1,Finkel2}
 by means of
the differential-difference operator technique.
It would be desirable to better understand
the relationship between our models and those in \cite{Finkel1,Finkel2}
(see also \cite{CC}).
A relevant fact in this respect is that if $\theta$ is an involution of a classical
Lie algebra, then
the `twisted Weyl group' (\ref{B.1E})
contains a standard Weyl group of $B$-type as a subgroup (see Appendix B).
It appears worthwhile to enquire about the use of the
twisted Weyl groups in constructing Dunkl type operators.
Second,  it is a natural task to describe the full family of spin Sutherland models,
and in particular all spinless special cases,
 that can be associated with  the  Hermann actions (\ref{3.2}).
These include also non-compact Riemannian manifolds
as starting points, and  the scattering theory of
the associated many-body models with spin appears worth investigating.
Third, the assumptions made in Chapter 2 can be weakened and modified in various
ways.
For example, one should extend the formalism to pseudo-Riemannian manifolds, such as
semisimple group manifolds of non-compact type (see \cite{FP_LMP}).
The strong conditions that we required in the definition of a `section' for convenience
should  be relaxed, as is done in some works in the theory of polar actions,
since there exist  examples for which these assumptions are not valid.
Finally, it is a challenge to understand whether an
infinite dimensional analogue of the formalism of polar actions can play a role
in describing Calogero type models with elliptic interaction potentials.

\medskip
\bigskip
\noindent{\bf Acknowledgements.}
We thank J. Balog for useful comments on the manuscript.
B.G.P. is also grateful to J. Harnad for hospitality in Montr\'eal.
The work of L.F. was supported in part by the Hungarian
Scientific Research Fund (OTKA grant
 T049495)  and by the EU network `ENIGMA'
(contract number MRTN-CT-2004-5652).

\renewcommand{\theequation}{\arabic{section}.\arabic{equation}}
\renewcommand{\thesection}{\Alph{section}}
\setcounter{section}{0}

\section{On the  `twisted measure factor' in  equation (\ref{3.28})}
\setcounter{equation}{0}
\renewcommand{\theequation}{A.\arabic{equation}}

In this appendix we explain that the second term
(the so-called `measure factor') appearing in the formula (\ref{2.17})
of the reduced Laplace-Beltrami operator gives just a \emph{constant}
for the twisted spin Sutherland models as stated by equation (\ref{3.28}).
In the non-twisted cases
 the analogous result is well-known \cite{Helg,OP}, and it turns out
that the `twisted measure factors' can be traced back to the non-twisted ones.
The density function associated with twisted conjugations
was also calculated in \cite{W}, and our arguments below follow from those in this reference.

Keeping the notations introduced in Chapter 3,
consider the real Cartan subalgebra $\cH_\rr$ of a complex simple Lie algebra $\cA$
with scalar product $\langle\ ,\  \rangle$ and associated root system $\Phi$.
Denote by $\Delta_{\cH_\rr}$ the Laplace-Beltrami operator of $(\cH_\rr, \langle\ ,\ \rangle)$,
using the restricted  scalar product\footnote{
In coordinates $q^i$ of $q\in \cH_\rr$ with respect to an orthonormal basis of $\cH_\rr$,
$\Delta_{\cH_{\rr}}= \sum_i \partial_i^2$, similarly to (\ref{3.27}).}.
Define
$\boldsymbol{\varrho}_\cG := \frac{1}{2} \sum_{\alpha\in \Phi_+} \alpha$,
where $\cG$ is the compact real form and
$\boldsymbol{\varrho}_\cG=\boldsymbol{\varrho}_\cG^{\mathrm{id}}$ in the
notation of (\ref{3.29}).
Introduce the standard  `density function' $\delta^{\frac{1}{2}}_{\Phi_+}: \cH_\rr \to \bR$ by
\be
\delta^{\frac{1}{2}}_{\Phi_+}(q) :=
\prod_{\alpha\in \Phi_+} \sin\left(\frac{\alpha(q)}{2}\right).
\label{A.1}\ee
It is well-known \cite{Helg,OP} that the following identity holds,
\be
\Delta_{\cH_\rr} \delta^{\frac{1}{2}}_{\Phi_+}= - \langle \boldsymbol{\varrho}_\cG,
\boldsymbol{\varrho}_\cG\rangle \delta^{\frac{1}{2}}_{\Phi_+}.
\label{A.2}\ee
This identity is independent of the  normalization of $\langle\ ,\ \rangle$, and the
value of the constant can be found from Freudenthal's strange formula
\be
g_{\cG} \dim(\cG)=12
\langle \boldsymbol{\varrho}_{\cG}, \boldsymbol{\varrho}_{\cG}\rangle_0,
\label{A.3}\ee
where $g_{\cG}$ is the dual Coxeter number of $\cG$  and
$\langle\ ,\ \rangle_0$ is the scalar product normalized so that
the length of the long roots is  $\sqrt{2}$.
Since the expressions in (\ref{3.25}) and (A.1) are equal if $\theta=\mathrm{id}$,
(A.2) yields the measure factor for the non-twisted spin Sutherland models.

To handle the twisted cases based on the
non-trivial involutive diagram automorphisms of the classical
Lie algebras, one may show  by case-by-case inspection that the `twisted density function'
(\ref{3.25}) satisfies
\be
\delta^\half(e^{\ri q})=
\prod_{\alpha \in {\mathfrak{R}}_+} \sin\left(\frac{\alpha(q)}{2}\right)
\prod_{\lambda \in {\mathfrak{W}}_+} \cos\left(\frac{\lambda(q)}{2}\right)
= {\cal C} \prod_{\varphi \in P_+}
\sin\left(\frac{\varphi(q)}{2}\right),
\label{A.4}\ee
where ${\cal C}$ is a constant and the set $P_+$
is given in terms of a corresponding root systems as follows:
\be
P_+ = \left \{
\begin{array}{cl}
\Phi_+(C_n) & \mbox{ if }\,\, \cA = D_{n + 1},\\
2 \Phi_+(B_n) & \mbox{ if }\,\, \cA = A_{2 n - 1},\\
2 \Phi_+(C_n) & \mbox{ if }\,\, \cA = A_{2 n}.
\end{array}
\right.
\label{A.5}\ee
This formula, which
is equivalent to the
formula of the density function given in \cite{W},
 is obtained from (\ref{3.25}) by using that
$\sin(2 x) = 2 \sin(x) \cos(x)$.
Referring to (\ref{3.29}) and  $P_+$ in (\ref{A.5}), it is also readily verified that
\be
{\boldsymbol{\varrho}}_\cG^\theta
= \half \sum_{\alpha \in {\mathfrak{R}}_+} \alpha +
 \half \sum_{\lambda \in {\mathfrak{W}}_+} \lambda
 =\half \sum_{\varphi \in P_+} \varphi.
\label{A.6}\ee
Equations (\ref{A.4}) and (\ref{A.5})  show that the twisted density function can be
cast into a non-twisted form. Since  the non-twisted
measure factors are  constants,
one may conclude that the twisted measure factors are constants, too.
As $P_+$ appears both in (\ref{A.4}) and in (\ref{A.6}), one obtains the
identity (\ref{3.28}) from (\ref{A.2}) applied to $(\cA^+, \cH_\rr^+)$.
With the scalar product (\ref{5.2}),
in the twisted $\cG=su(N)$ cases the values of the constants
on the right-hand side of (\ref{3.28}) are  found from the relations
\be
6\langle \boldsymbol{\varrho}^\theta_\cG,
\boldsymbol{\varrho}^\theta_\cG \rangle
= \left \{
\begin{array}{cl}
n (2 n - 1) (2 n + 1)
& \mbox{ if }\,\, \cG = su(2n),\\
2 n (n + 1) (2 n + 1)
& \mbox{ if }\,\, \cG = su(2 n+1).
\end{array}
\right.
\label{A.7}\ee
These relations can be checked  by explicit calculation or by application of the strange formula
(\ref{A.3}) taking care of the normalizations involved.

The measure factors are constants for all diagram automorphisms of $E_6$ and $D_4$ as well.
In fact, the non-twisted form of the density function is established in \cite{W}
in these cases, too.

\section{Generalized Weyl groups for twisted conjugations}
\setcounter{equation}{0}
\renewcommand{\theequation}{B.\arabic{equation}}

For completeness, we here recall from \cite{W,M,MW} the structure of the
generalized Weyl group associated with the twisted conjugation action (\ref{3.12}).
 As in the main text, we assume that $G$ is simply connected and $\Theta$ is
 involutive.

In this case (\ref{2.18}) becomes the `twisted Weyl group'
$W(G,T^\Theta, \Theta)$ in (\ref{B.1E}).
Let $N_G(T)$ denote the normalizer of $T$ in $G$ defined in the
usual manner, i.e.~$N_G(T^\Theta, I^\Theta)$ in (\ref{B.2E}) yields
$N_G(T)$ for $\Theta=\mathrm{id}$,
and consider also $N_{G^\Theta}(T^\Theta)$, where $G^\Theta$ is the (connected)
fixed point subgroup of $\Theta$ in $G$.
According to \cite{M}, the twisted Weyl group is given by the  semidirect product
\be
W(G,T^\Theta,\Theta) \cong W(G^\Theta, T^\Theta) \ltimes (T/T^\Theta)^\Theta.
\label{B.1}\ee
Here $W(G^\Theta, T^\Theta)=N_{G^\Theta}(T^\Theta)/T^\Theta$ and the finite Abelian group
$(T/T^\Theta)^\Theta$  is formed by the fixed points of the induced action of $\Theta$
on $(T/T^\Theta)$.
Note that $W(G^\Theta, T^\Theta)$ acts  by automorphisms of $(T/T^\Theta)^\Theta$
naturally as
\be
(g T^\Theta).(tT^\Theta) = g t g^{-1} T^\Theta
\quad\hbox{for any}\quad
gT^\Theta \in W(G^\Theta, T^\Theta),\,\,
tT^\Theta \in (T/T^\Theta)^\Theta.
\label{B.2}\ee
This is well-defined since
$N_{G^\Theta}(T^\Theta) < N_G(T)$,
which follows from the  fact \cite{Kac} that $T^\Theta$ contains a regular element of
$G$ (whose centralizer is $T$).
As induced from the $I^\Theta$ action (\ref{3.12}),
the subgroup $W(G^\Theta,T^\Theta)$ in (\ref{B.1}) acts on  $T^\Theta$ by
its ordinary Weyl group action.
The map $S: (T/T^\Theta)^\Theta \to T^\Theta$ defined by
\be
S: t T^\Theta \mapsto S(tT^\Theta):= \Theta(t) t^{-1}
\label{B.3}\ee
is an injective homomorphism, and the normal subgroup $(T/T^\Theta)^\Theta$ in (\ref{B.1})
acts on the `section' $T^\Theta$ by the translations  generated by its image in $T^\Theta$:
\be
(tT^\Theta). t_0 = S(t T^\Theta) t_0
\quad\hbox{for any}\quad
 tT^\Theta\in (T/T^\Theta)^\Theta,\,\, t_0 \in T^\Theta.
\label{B.4}\ee
Correspondingly,  (\ref{B.1}) can be rewritten in the form
\be
W(G,T^\Theta,\Theta)\cong W(G^\Theta,T^\Theta) \ltimes S((T/T^\Theta)^\Theta),
\label{B.5}\ee
where $W(G^\Theta,T^\Theta)$ acts naturally on the finite subgroup
$S((T/T^\Theta)^\Theta) < T^\Theta$.

To proceed further, let us introduce the lattices
\be
\cL := \operatorname{Ker}\left(\exp|_{\cT} \right):= \{ X\in \cT \,\vert\, \exp(X) =e\} < \cT
\quad\hbox{and}\quad
\cL^\theta:= \cL \cap \cT^\theta,
\label{B.6}\ee
where now $\cT^\theta$  denotes the fixed point set of $\theta$ in $\cT$.
Note that $T\cong \cT /\cL$
and $T^\Theta \cong \cT^\theta/\cL^\theta$.
Let $p: \cT \to \cT^\theta$ be the orthogonal projection with respect to the Killing form.
Then \cite{MW} one has yet another natural isomorphism:
\be
S((T/T^\Theta)^\Theta) \cong p(\cL)/\cL^\theta.
\label{B.7}\ee
To understand this, first notice that
if $\theta(X_-)=-X_-$, then
$\Theta(\exp(\frac{1}{2} X_-) T^\Theta)= \exp(\frac{1}{2} X_-) T^\Theta$ is equivalent to
$\exp(-X_-)\in T^\Theta$, which requires that $\exp(-X_-)=\exp(X_+)$ for some
$X_+\in \cT^\theta$.
Next, if
$X = (X_+ + X_-) \in \cL$ with $X_+=p(X)\in \cL^\theta$, then
$\exp(\frac{1}{2} X_-) T^\Theta \in (T/T^\Theta)^\Theta$
and
\be
S(\exp(\frac{1}{2} X_-
) T^\Theta )=\exp(-X_-) = \exp(X_+)\in p(\cL)/\cL^\theta < T^\Theta \cong \cT^\theta/\cL^\theta.
\label{B.8}\ee
Every element of $p(\cL)/\cL^\theta$ corresponds by (\ref{B.8})  to a unique
element of $S((T/T^\Theta)^\Theta)$.
Taking into account (\ref{B.7}),
the semidirect product
\be
W(G,T^\Theta, \Theta)\cong
W(G^\Theta,T^\Theta) \ltimes p(\cL)/\cL^\theta
\label{B.9}\ee
can be `lifted' to yield the infinite group
\be
W(G^\Theta,T^\Theta) \ltimes p(\cL),
\label{B.10}\ee
which can be identified \cite{W,MW} as
the Weyl group of the twisted affine Lie algebra associated
with the pair $(\cA, \theta)$.
The twisted conjugacy classes in $G$ are parametrized by the orbits of the twisted
Weyl group in the section $T^\Theta$, i.e., they form the (stratified) space
\be
[\cT^\theta/\cL^\theta]
/ [W(G^\Theta,T^\Theta)\ltimes p(\cL)/\cL^\theta]
  \cong \cT^\theta/ [W(G^\Theta,T^\Theta)\ltimes p(\cL)].
\label{B.11}\ee
The set on the right hand side of (\ref{B.11}) is the space of orbits for
the twisted affine Weyl group acting on $\cT^\theta$, which is  described in \cite{Kac}.
By using this, Mohrdieck and Wendt \cite{MW} gave parametrizations for
the twisted conjugacy classes of any simply connected, connected $G$.
In our cases of interest,
this is equivalently given by the domains
listed at the end of Chapter 3.

Let us display the twisted Weyl groups in the picture
(\ref{B.5}) for $G=SU(N)$.
The fixed point sets $T^\Theta_N$ in the maximal torus $T_N$ of $SU(N)$ are
\bea
&& \quad T_{2n}^\Theta =
\{ t\,\vert\, t=\operatorname{diag}(t_1,\ldots, t_n, t_n^{-1},\ldots, t_1^{-1}),\,\,
\vert t_i \vert =1,\,\, 1\leq i \leq n\}, \nonumber\\
&& T_{2n+1}^\Theta = \{ t\,\vert\, t=\operatorname{diag}(t_1,\ldots, t_n,1, t_n^{-1},\ldots, t_1^{-1}),\,\,
\vert t_i \vert =1,\,\, 1\leq i \leq n\}.
\label{B.12}\eea
In both cases
\be
W(SU(N)^\Theta,T_N^\Theta)\cong W_{B_n} \cong S_n \ltimes (\bZ_2)^{\times n}.
\label{B.13}\ee
The symmetric group $S_n$  permutes the components $t_1,\ldots, t_n$ of $t\in T_N^\Theta$
and, for any fixed $1\leq i\leq n$, the generator $\tau_i \in (\bZ_2)^{\times n} $
exchanges the components $t_i$ and $t_i^{-1}$ of $t$.
It is easy to compute that $S((T_{N}/T_{N}^\Theta)^\Theta)$ consists of the elements
$\sigma$ of the form
\bea
&&\sigma= \operatorname{diag}(\sigma_1,\ldots, \sigma_n, \sigma_n,\ldots, \sigma_1)
\quad\hbox{with}\quad
\sigma_i\in \{ \pm 1\},\,\, \prod_{i=1}^n \sigma_i =1\quad\hbox{if}\quad N=2n,
\nonumber\\
&&\sigma= \operatorname{diag}(\sigma_1,\ldots, \sigma_n, 1, \sigma_n,\ldots, \sigma_1)
\quad\hbox{with}\quad
\sigma_i\in \{ \pm 1\}\quad\hbox{if}\quad N=(2n+1).
\label{B.14}\eea
Thus we see that
\bea
&&W(SU(2n)^\Theta,T_{2n}^\Theta,\Theta) \cong W_{B_n} \ltimes (\bZ_2)^{\times (n-1)},
\nonumber\\
&& W(SU(2n+1)^\Theta,T_{2n+1}^\Theta,\Theta) \cong W_{B_n} \ltimes (\bZ_2)^{\times n}.
\label{B.15}\eea
The $S_n$ subgroup of $W_{B_n}$ acts by permuting the components
$\sigma_i$ of $\sigma \in S((T_N/T_N^\Theta)^\Theta)$, while the generators $\tau_i$
act trivially on  $S((T_N/T_N^\Theta)^\Theta)$.

Finally, let us remark that one may introduce (physically and mathematically different)
`extended versions'
of the twisted spin Sutherland models of Proposition 3.1 by postulating the dense open subset
$\left(\exp|_{\cT}\right)^{-1}(\hat T^\Theta) \subset \cT^\theta$ to be the
configuration space, instead of the alcove $\check \cT^\theta$.
For suitable domains of the formal Hamiltonian obtained from (\ref{3.30}),
these extended models possess
the twisted affine Weyl group (\ref{B.10}) as a symmetry group.

\section{On the derivation of Propositions 5.1 and 5.2}
\setcounter{equation}{0}
\renewcommand{\theequation}{C.\arabic{equation}}

We here sketch how we obtained
the explicit formulae (\ref{5.19}) and (\ref{5.20}) for the
oscillator realizations of the reduced Hamiltonian operators
in the twisted $SU(N)$ cases.
By applying the conventions of Chapter 3, we first note that both for
$\varphi \in \mathfrak{R}_+$ and for
$\varphi \in \mathfrak{W}_+$ we have
\be
\rho'(Y^\pm_\varphi)^2 +\rho'(Z^\pm_\varphi)^2
= -
\rho'(X^\pm_\varphi) \rho'(X^\pm_{-\varphi})
- \rho'(X^\pm_{-\varphi}) \rho'(X^\pm_\varphi).
\label{C.1}\ee
To spell out   $\Delta_{\mathrm{red}}$ (\ref{3.30})
in terms of the oscillators in (\ref{5.6}), we need matrix realizations for
the generators $X^\pm_\varphi$ and $K^-_j$.
The involution $\theta$ (\ref{5.11})
preserves the Cartan subalgebra $\cH$ spanned by diagonal
matrices, while the root and weight vectors in (\ref{C.1}) are
off-diagonal matrices.
It is enough to list the latter base elements for positive $\varphi$, since
$X_{-\varphi}^\pm = (X_{\varphi}^\pm)^T$ holds.

When $N$ is even, $N = 2 n$,
$\cH^+_\rr$ consists of the elements (\ref{3.34})
and in $\cH^-_\rr$ we define an orthonormal basis
$\{ K^-_j \}_{j = 1}^{n - 1}$ by the matrices
\be
K^-_j
:= \frac{1}{\sqrt{2 j (j + 1)}}
\left(
\sum_{i = 1}^j (E_{i, i} + E_{2 n + 1 - i, 2 n + 1 - i})
- j (E_{j + 1, j + 1} + E_{2 n - j, 2 n - j})
\right).
\label{C.2}\ee
As one can easily verify, the following off-diagonal matrices
\bea
&& X^+_{2 e_k} := E_{k, 2 n + 1 -k}
\quad
(1 \leq k \leq n),
\nonumber\\
&& X^\pm_{e_k - e_l}
:= \frac{1}{\sqrt{2}} \left(
E_{k, l} \mp (-1)^{k + l} E_{2 n + 1 -l, 2 n + 1 -k}
\right)
\quad
(1 \leq k < l \leq n),
\nonumber\\
&& X^\pm_{e_k + e_l}
:= \frac{1}{\sqrt{2}} \left(
E_{k, 2 n + 1 - l} \pm (-1)^{k + l} E_{l, 2 n + 1 -k}
\right)
\quad
(1 \leq k < l \leq n)
\label{C.3}\eea
provide root and weight vectors satisfying the conventions of subsection 3.1,
using the scalar product (\ref{5.2}).

When $N$ is odd, $N = 2 n + 1$, $\cH_\rr^+$ is parametrized
according to (\ref{3.37}) and in
$\cH^-_\rr$ we introduce dual bases $\{ L_j \}_{j = 1}^n$,
$\{ L^j \}_{j = 1}^n$ defined by
\be
L_j := E_{j, j} + E_{2 n + 2 - j, 2 n + 2 - j} - 2 E_{n + 1, n + 1},
\quad
L^j := \frac{1}{2} L_j - \frac{1}{2 n + 1} \sum_{i = 1}^n L_i.
\label{C.4}\ee
These permit to evaluate the third term of (\ref{3.30}) by means of the relation
\be
\sum_j \rho'(\ri K^-_j)^2 = - \sum_j \rho'(L^j) \rho'(L_j).
\ee
The normalized root and weight vectors
in $sl(2 n + 1, \bC)$ are furnished by the matrices
\bea
&& X^-_{2 e_k} := E_{k, 2 n + 2 -k}
\quad
(1 \leq k \leq n),
\nonumber\\
&& X^\pm_{e_k} := \frac{1}{\sqrt{2}} \left(
E_{k, n + 1} \pm (-1)^{k + n} E_{n + 1, 2 n + 2 - k} \right)
\quad
(1 \leq k \leq n),
\nonumber\\
&& X^\pm_{e_k - e_l}
:= \frac{1}{\sqrt{2}} \left(
E_{k, l} \mp (-1)^{k + l} E_{2 n + 2 - l, 2 n + 2 - k} \right)
\quad
(1 \leq k < l \leq n),
\nonumber\\
&& X^\pm_{e_k + e_l}
:= \frac{1}{\sqrt{2}} \left(
E_{k, 2 n + 2 - l} \mp (-1)^{k + l} E_{l, 2 n + 2 - k} \right)
\quad
(1 \leq k < l \leq n).
\label{C.5}\eea

Based on the above introduced generators and equation (\ref{5.6}), the
explicit formulae displayed in Propositions 5.1 and 5.2
follow by straightforward calculation.


\begin{thebibliography}{99}

\bibitem{Cal}
F. Calogero,
{\it Solution of the one-dimensional $N$-body problem with quadratic and/or
inversely quadratic pair potentials},
J. Math. Phys.  \textbf{12} (1971) 419-436

\bibitem{Sut}
B. Sutherland,
{\it Exact results for a quantum many body problem in one dimension},
Phys. Rev. \textbf{ A4} (1971) 2019-2021

\bibitem{OPRep}
M.A. Olshanetsky and A.M. Perelomov,
{\it Quantum integrable systems related to Lie algebras},
Phys. Rept.  \textbf{94} (1983) 313-404

\bibitem{Heck}
G. Heckman,
{\it Hypergeometric and spherical functions},
pp. 1-89 in:
G. Heckman and H. Schlichtkrull,
Harmonic Analysis and Special Functions on Symmetric Spaces,
Perspectives in Mathematics 16, Academic Press, 1994

\bibitem{Dunkl}
C.F. Dunkl and Y. Xu,
Orthogonal Polynomials of Several Variables,
Encyclopaedia of Mathematics and its Applications 81, Cambridge University Press,
2001

\bibitem{SutWSci}
B. Sutherland,
Beautiful Models, World Scientific, 2004

\bibitem{Cher}
I. Cherednik,
Double Affine Hecke Algebras,
 London Mathematical Society Lecture Notes Series 319,
Cambridge University Press, 2005

\bibitem{Sas}
R. Sasaki,
Quantum Calogero-Moser Systems,
pp. 123-129 in: Encyclopaedia of Mathematical Physics,
Academic Press, 2006

\bibitem{Eting}
P. Etingof,
Calogero-Moser Systems and Representation Theory,
European Mathemathical Society, Z\"urich, 2007,
{\tt arXiv:math/0606233 [math.QA]}

\bibitem{Poly}
A.P. Polychronakos,
{\it Physics and mathematics of Calogero particles},
J. Phys.  \textbf{A39} (2006) 12793-12845,
{\tt arXiv:hep-th/0607033}

\bibitem{FP_polar} L. Feh\'er and B. G. Pusztai,
{\it Hamiltonian reductions of free particles
under polar actions of compact Lie groups},
{\tt arXiv:0705.1998 [math-ph]} (to appear in Theor. Math. Phys.)

\bibitem{FP_ROMP} L. Feh\'er and B. G. Pusztai,
{\it On the self-adjointness of certain reduced Laplace-Beltrami operators},
{\tt arXiv:0707.2708 [math-ph]} (to appear in Rep. Math. Phys.).

\bibitem{PT}
R. Palais and C.-L. Terng,
{\it A general theory of canonical forms},
Trans. Amer. Math. Soc.  \textbf{300} (1987) 771-789

\bibitem{OP}
M.A. Olshanetsky and A.M. Perelomov,
{\it Quantum systems related to root systems, and radial parts of Laplace operators},
Funct. Anal. Appl.  \textbf{12}  (1978) 121-128,
{\tt arXiv:math-ph/0203031}

\bibitem{EFK}
P.I. Etingof, I.B. Frenkel and A.A. Kirillov Jr.,
{\it Spherical functions on affine  Lie groups},
Duke Math. J.  \textbf{80} (1995) 59-90,
{\tt arXiv:hep-th/9407047}

\bibitem{Helg}
S. Helgason, Groups and Geometric Analysis, Academic Press, 1984

\bibitem{Obl}
A. Oblomkov,
{\it Heckman-Opdam's Jacobi polynomials for the
 $BC_n$ root system and generalized spherical functions},
 Adv. Math. \textbf{186} (2004) 153-180,
 {\tt arXiv:math/0202076 [math.RT]}

\bibitem{FP_LMP}
L. Feh\'er and B.G. Pusztai,
{\it A class of Calogero type reductions of free motion on a simple Lie group},
Lett. Math. Phys.  \textbf{79} (2007) 263-277,
{\tt arXiv:math-ph/0609085}

\bibitem{Hoch}
S. Hochgerner,
{\it Singular cotangent bundle reduction and spin Calogero-Moser systems},
{\tt arXiv:math/0411068 [math.SG]} (to appear in Diff. Geom. Appl.)

\bibitem{FP1}
L. Feh\'er and B.G. Pusztai,
{\it Spin Calogero models obtained from dynamical r-matrices and geodesic motion},
Nucl. Phys.  \textbf{B734} (2006) 304-325,
{\tt arXiv:math-ph/0507062}

\bibitem{FP2}
L. Feh\'er and B.G. Pusztai,
{\it Spin Calogero models associated with Riemannian symmetric spaces of negative curvature},
Nucl. Phys.  \textbf{B751}  (2006) 436-458,
{\tt arXiv:math-ph/0604073}

\bibitem{W}
R. Wendt,
{\it Weyl's character formula for non-connected Lie groups and orbital theory for twisted
affine Lie algebras},
J. Funct. Anal.  \textbf{180} (2001) 31-65,
{\tt arXiv:math/9909059 [math.RT]}

\bibitem{M}
S. Mohrdieck,
Conjugacy Classes of Non-Connected Semi-Simple Algebraic Groups, PhD thesis,
University of Hamburg, 2000

\bibitem{MW}
S. Mohrdieck and R. Wendt,
{\it Integral conjugacy classes of compact Lie groups},
Manuscripta Math. \textbf{114} (2004) 531-547,
{\tt arXiv:math/0303118 [math.QA]}

\bibitem{IT} S. Tanimura and T. Iwai, {\it Reduction of quantum systems on Riemannian
manifolds with symmetry and application to molecular mechanics},
J. Math. Phys. \textbf{41} (2000) 1814-1842,
{\tt arXiv:math-ph/9907005}

\bibitem{Lands} N.P. Landsman,
Mathematical Topics Between Classical and Quantum Mechanics,
Springer, 1998

\bibitem{Davis}
M. Davis,
{\it Smooth $G$-manifolds as collections of fiber bundles},
Pacific J. Math.  \textbf{77} (1978) 315-363

\bibitem{GOV}
V.V. Gorbatsevich, A.L. Onishchik and E.B. Vinberg,
Foundations of Lie Theory and Lie Transformation Groups,
Springer, 1997

\bibitem{HPTT}
E. Heintze, R. Palais, C.-L. Terng and G. Thorbergsson,
{\it Hyperpolar actions on symmetric spaces},
in: Geometry, topology and physics for Raoul Bott, ed. S.-T. Yau, International Press, 1994

\bibitem{Kol}
A. Kollross,
{\it Polar actions on symmetric spaces},
{\tt arXiv:math/0506312 [math.DG]} and references therein

\bibitem{Kac} V.G. Kac,  Infinite Dimensional Lie Algebras,
third edition, Cambridge University Press, 1990

\bibitem{AM}
A. Alekseev and E. Meinrenken,
{\it Clifford algebras and the classical dynamical Yang-Baxter equation},
Math. Res. Lett.  \textbf{10} (2003) 253-268,
{\tt arXiv:math/0209347  [math.RT]}

\bibitem{Knapp}
A.W. Knapp,
Lie Groups Beyond an Introduction, Progress in Mathematics 140, Birkh\"auser, 2002

\bibitem{Berezin}
F.A. Berezin,
{\it Some remarks about the associated envelope of a Lie algebra},
Funct. Anal. Appl.  \textbf{1} (1967) 91-102

\bibitem{FH}
W. Fulton and J. Harris,
Representation Theory, A First Course, Springer, 1991

\bibitem{Finkel1}
F. Finkel, D. G\'omez-Ullate, A. Gonz\'alez-L\'opez, M.A. Rodr\'\i guez and R. Zhdanov,
{\it New spin Calogero-Sutherland models related to $B_N$-type Dunkl operators},
 Nucl. Phys. \textbf{B613} (2001) 472-496,
{\tt arXiv:hep-th/0103190}

\bibitem{Finkel2}
F. Finkel, D. G\'omez-Ullate, A. Gonz\'alez-L\'opez, M.A. Rodr\'\i guez and R. Zhdanov,
{\it On the Sutherland spin model of $B_N$ type and its associated spin chain},
Commun. Math. Phys.  \textbf{233} (2003) 191-209,
{\tt arXiv:hep-th/0202080}

\bibitem{CC}
V. Caudrelier and N. Crampe,
{\it Integrable $N$-particle Hamiltonians with Yangian or reflection algebra symmetry},
J. Phys. \textbf{A37} (2004) 6285-6298,
{\tt arXiv:math-ph/0310028}


\end{thebibliography}
\end{document}